\documentclass[manuscript,screen]{acmart}
\usepackage{listings}
\usepackage{color}
\usepackage[export]{adjustbox}
\usepackage{graphicx}

\usepackage{subfig}
\usepackage{tcolorbox}
\usepackage{caption}
\usepackage{array}
\usepackage{multirow}
\usepackage[normalem]{ulem}
\usepackage{xcolor}
\usepackage{amsmath}
\usepackage{algorithm}
\usepackage{algpseudocode}
\usepackage[normalem]{ulem}
\definecolor{Pgreen}{RGB}{0,166,147}

\AtBeginDocument{%
  \providecommand\BibTeX{{%
    \normalfont B\kern-0.5em{\scshape i\kern-0.25em b}\kern-0.8em\TeX}}}

\setcopyright{acmcopyright}
\copyrightyear{2023}
\acmYear{2023}
\acmDOI{XXXXXXX.XXXXXXX}

\begin{document}

\title[Probing Models Trained on Code]{DeepCodeProbe: Towards Understanding What Models Trained on Code Learn}

\author{Vahid Majdinasab}

\email{vahid.majdinasab@polymtl.ca}
\orcid{0000-0003-4411-0810} 
\affiliation{%
  \institution{Polytechnique Montreal}
  \streetaddress{}
  \city{Montreal}
  \state{Quebec}
  \country{Canada}
  \postcode{}
}

\author{Amin Nikanjam}
\orcid{0000-0002-0440-6839}
\affiliation{%
  \institution{Polytechnique Montreal}
  \streetaddress{}
  \city{Montreal}
  \state{Quebec}
  \country{Canada}
  \postcode{}}
\email{amin.nikanjam@polymtl.ca}

\author{Foutse Khomh}
\orcid{0000-0002-5704-4173}
\affiliation{\institution{Polytechnique Montreal}
  \streetaddress{}
  \city{Montreal}
  \state{Quebec}
  \country{Canada}
  \postcode{}
}
\email{foutse.khomh@polymtl.ca}
\newcommand{\dcp}{DeepCodeProbe }
\newcommand{\dcu}{$<d, c, u>$ }
\newcommand{\cu}{$<c, u>$ }

\renewcommand{\shortauthors}{Majdinasab et al.}
\
\begin{abstract}
    Machine Learning (ML) models trained on code and artifacts extracted from them (e.g., version control histories, code differences, etc.), provide invaluable assistance for software maintenance tasks. Despite their good performance, these models can make errors that are difficult to understand due to their large latent space and the complexity of interactions among their internal variables. The lack of interpretability in these models' decision-making processes raises concerns about their reliability, especially in safety-critical applications. Furthermore, the specific representations and features these models learn from their training data remain unclear, further contributing to the interpretability challenges and hesitancy in adopting these models in safety-critical contexts. To address these challenges and provide insights into what these models learn from code, we present DeepCodeProbe, a probing approach designed to investigate the syntax and representation learning capabilities of trained ML models aimed at software maintenance tasks. Applying DeepCodeProbe to state-of-the-art code clone detection, code summarization, and comment generation models provides empirical evidence that while small models capture abstract syntactic representations relevant to their tasks, their ability to learn the complete programming language syntax is limited. Our results show that increasing model capacity by increasing the number of parameters without changing the architecture improves syntax learning to an extent but introduces trade-offs in training and inference time alongside overfitting. \dcp also uncovers specific code patterns and representations the models learn based on how code is represented for training the models. Our proposed probing approach uses a novel abstract syntax tree-based data representation that allows for probing models for syntax information. Leveraging our experimentation with the different models, we also provide a set of best practices for training models on code, to enhance both the models' performance and interpretability for code-related tasks.\\ Additionally, our open-source replication package allows the application of \dcp to interpret other code-related models. Our work addresses the critical need for reliable and trustworthy ML models in software maintenance. The insights from \dcp can guide the development of more robust and interpretable models tailored for this domain.
    
\end{abstract}



\received{July 2024}

\maketitle

\section{Introduction}\label{sec:introduction}
    Effective software maintenance is key to ensuring the long-term reliability and sustainability of software systems. Over time, growing codebases and rising code complexity pose challenges for maintaining software systems \cite{koch2005evolution}. Machine learning (ML) models offer assistance in addressing software maintenance tasks such as code clone detection, fault localization, code smell detection, etc. \cite{yang2022survey}. Utilizing ML models trained on expansive code and related artifacts (e.g., statistics about the code such as the number of defined variables and function calls, documentation, control/data flow graphs, etc.) allows for facilitating various phases of software maintenance \cite{yang2022survey}. For example, data from the revision history of a project can be used for defect prediction, helping to address problems early on~\cite{barbez2019deep}. Moreover, leveraging Natural Language Processing (NLP) techniques alongside ML models has been shown to improve documentation interpretation, generation, and change assessment; reducing developers' time and mistakes \cite{hu2018deep}. Furthermore, Applying ML models for code clone detection helps to locate similar code inside the codebase, assisting developers in handling issues affecting multiple code copies \cite{lei2022deep}.

Despite these benefits, however, there exist challenges concerning model interpretability \cite{rudin2019stop} and the reliability of ML models in production \cite{sculley2015hidden}. Due to the black-box nature of state-of-the-art ML models, interpreting their decisions is difficult \cite{carvalho2019machine}. This lack of interpretability can lead to significant issues in software development and maintenance. For instance, when developers cannot interpret the decisions made by code models, debugging and repairing code becomes more complex and time-consuming \cite{tang2024study, zhou2023concerns}. Misinterpreted model outputs can introduce subtle bugs that are hard to trace, potentially compromising software reliability and increasing maintenance costs \cite{majdinasab2023assessing, pan2024lost, tambon2024bugs}. Additionally, in collaborative environments, the inability to understand model decisions can hinder effective team communication and slow down the development process \cite{rasnayaka2024empirical,nguyen2024beginning,wang2024rocks}. Therefore, understanding what ML models learn from their training data and how they generate outputs regarding the underlying task, is crucial for establishing their reliability and trustworthiness within safety-critical applications and production environments \cite{hall2019proposed, ashoori2019ai}. Furthermore, examining the internals of ML models contributes to improving the overall performance by identifying decision-making errors that are amenable to fine-tuning or alternative architectural and data representation choices of models. One of the approaches used for this purpose is probing; in which a simple classifier is trained on the embeddings extracted from a model to predict a property of interest \cite{karmakar2021pre}. Probing ML models facilitates understanding the relationships between the models' input features and output predictions, allowing practitioners to assess whether the learned representations are aligned with expectations \cite{zhao2020safety}. Given the recent rise of Large Language Models (LLM), researchers have been proposing novel approaches for probing LLMs for a variety of tasks and training data \cite{choudhary2022interpretation}. However, the majority of the proposed probing approaches are only applicable to transformer-based models with a large number of parameters. As such, there exists a gap in the current literature with a lack of probing techniques focusing on much smaller models trained on code for software maintenance tasks. To address this gap and given the benefits of having reliable models for software maintenance, we propose DeepCodeProbe, a probing approach designed for ML models trained on code for software maintenance tasks. \dcp allows gaining insights into what ML models learn from their training data. Our probing approach is focused on identifying whether ML models used in software maintenance tasks learn the syntax of the programming languages that they have been trained on and the representations they learn from their inputs to achieve their objectives. 

\dcp is built upon the principles of AST-Probe \cite{hernandez2022ast}, which examines LLMs' capability of encoding programming languages' syntax into their latent space. However in contrast to AST-Probe, \dcp specifically targets smaller ML models designed for software maintenance, distinct from large, transformer-based models. \dcp is based on a novel data representation and embedding extraction approach for probing models trained on code. Leveraging DeepCodeProbe, we examine how models trained on code capture programming language syntax within their latent spaces and assess the effectiveness of our data representation and embedding extraction. Furthermore, we employ our probing approach to understand what representations of data the models under study learn from their training data to achieve their objectives. Finally, we formulate a set of best practices for training ML models on code for software maintenance tasks.

Briefly, this paper makes the following contributions:
\begin{itemize}
    \item We propose a new probing approach for determining whether ML models trained on code learn the syntax of the programming language.
    \item We use \dcp to investigate how ML models represent information from code in their latent space. We also investigate the effect of model size on their syntax learning capabilities.
    \item We formulate a set of best practices for training ML models for software maintenance.
    \item We publish all our data and scripts to allow for the replication and use of our approach at \cite{rep_package} for other researchers to use.
\end{itemize}

The rest of the paper is organized as follows: In Section \ref{sec:background}, we present the background concepts and the related literature. We introduce \dcp and our methodology alongside the details of our probing approach in Section \ref{sec:dcp}. We describe our experiment design alongside the models under study in Section \ref{sec:exp_design}. We present our results in Section \ref{sec:results}. We review the related works in Section \ref{sec:related_works}, present the best practices for training small models on code in Section \ref{sec:discussion}, and discuss threats to the validity of our study in Section \ref{sec:threats_to_validity}. Finally, we conclude the paper in Section \ref{sec:conclusion} and outline some avenues for future works.

\section{Background}\label{sec:background}
    In this section, we will review the necessary background and concepts related to our probing approach. \dcp is designed to probe code models for software maintenance, therefore, we will first review the literature on ML for software maintenance. Afterward, we will discuss the probing approaches that have been proposed in NLP for ML models. Finally, as \dcp is built on top of AST-Probe, we will describe AST-Probe and its underlying concepts.

\subsection{Machine Learning for Software Maintenance} \label{subsec:background:machine_learning_for_software_maintenance}
Software maintenance is a critical activity throughout the software development lifecycle. It involves continuous updates, testing, and improvements to ensure that software functions optimally according to evolving user requirements while preserving its original capabilities \cite{rajlich2014software}. Given the increasing sophistication and size of contemporary software systems \cite{koch2005evolution}, their maintenance can become laborious, time-consuming, and error-prone \cite{mens2005challenges, benestad2010understanding}. Consequently, numerous ML-based techniques have been proposed to simplify and enhance various aspects of software maintenance.
    
By training ML models on different software artifacts, such as source code \cite{gupta2017deepfix, bhatia2018neuro}, version control history \cite{barbez2019deep}, documentation \cite{gros2020code}, and software metrics \cite{kumar2016hybrid}, these models can assist developers at various stages of software maintenance \cite{yang2022survey}. The primary applications of ML in software maintenance include:
    
    \begin{itemize}
        \item Defect prediction: ML models have been proposed to help developers anticipate potential problems in specific areas of the codebase. This enables efficient resource allocation and proactive issue resolution \cite{tong2018software, dam2018deep, liu2018connecting}.
        
        \item Program repair: ML models have been proposed to help with automatically fixing common bugs. This can significantly reduce manual effort, minimize human errors, and save resources \cite{bhatia2018neuro, tufano2019empirical, lutellier2020coconut}.
        
        \item Code clone detection: ML models have been proposed to detect clones in the codebase by identifying duplicate or similar segments of codes to facilitate refactoring efforts, ensuring consistency and reducing redundancy \cite{white2016deep, gao2019teccd}.
        
        \item Code summarization: Proposed ML models allow for understanding complex code logic, easier maintainability, and collaboration by generating abstract representations of code blocks \cite{leclair2020improved, ahmad2020transformer}.

        \item Comment generation: By producing descriptive comments,  ML models allow for automatically increasing the intelligibility of code, ease of knowledge transfer, and support long-term maintenance activities \cite{hu2018deep, li2022setransformer}.
    \end{itemize}

    As our study is focused on ML models that are trained on code, we choose two representative tasks in software maintenance, namely code clone detection and code summarization, for probing and review the literature on these tasks, below.
    
    \subsubsection{Code Clone Detection} \label{subsec:background:code_clone_detection}
        Code Clone Detection (CCD) is an important activity in software maintenance as a fault in one project can also be present in other projects with similar code. Identifying and correcting these clones is essential to ensure overall software quality. %
        Tasks such as software refactoring, quality analysis, and code optimization often necessitate the identification of semantically and syntactically similar code segments \cite{roy2009comparison}. In the CCD literature, clones are categorized into four types \cite{roy2007survey}:

        \begin{itemize}
            \item \textit{Type-1:} Exact copies except for whitespace/comments.
            \item \textit{Type-2:} Syntactically identical with different identifiers.
            \item \textit{Type-3:} Similar fragments with added/removed statements.
            \item \textit{Type-4:} Different syntax but similar semantics.
        \end{itemize}
        
        Detecting Type-1 and Type-2 clones is relatively straightforward, as they can be identified through syntactic comparisons. However, Type-3 and Type-4 clones pose a greater challenge, as they require capturing semantic similarities beyond surface-level syntax. Traditional approaches, such as text-based or token-based comparisons, often struggle with these more complex clone types \cite{lei2022deep}. To address this problem, the representation learning capabilities of Deep Learning (DL) models allow for the use of representations extracted from code such as Abstract Syntax Trees (ASTs), Control Flow Graphs (CFGs), and software metrics for CCD. By utilizing such representations, DL-based approaches such as FCCA \cite{hua2020fcca}, CoCoNut \cite{lutellier2020coconut}, and Cclearner \cite{li2017cclearner} have achieved high performance on CCD benchmarks.  
    
    \subsubsection{Code Summarization and Comment Generation} \label{subsec:background:code_summarization_and_comment_generation}
        Code summarization and comment generation are closely related tasks that aim to enhance code comprehension and maintainability. Code summarization involves generating concise natural language descriptions that capture the functional behavior or high-level semantics of a given code snippet \cite{zhang2022survey}. On the other hand, comment generation focuses on producing comments that explain the purpose and functionality of specific code segments \cite{song2019survey}. Several DL-based approaches have been proposed for code summarization and comment generation. Allamanis et al. \cite{allamanis2016convolutional} train a Convolutional Neural Network (CNN) with an attention mechanism for code summarization. This approach leverages the strengths of CNNs in capturing local patterns while the attention mechanism helps in focusing on the most relevant parts of the code for generating accurate summaries. Another work done by Hu et al. \cite{hu2018deep} uses the ASTs constructed from code snippets to train a sequence-to-sequence model for comment generation for Java methods. By encoding the structural information from the AST, their model can better capture the code's semantics and produce more relevant comments tailored to the specific code segments. Wei et al. \cite{wei2019code} trained two models for code generation and code summarization by initializing a sequence-to-sequence model for each task and training them alongside each other by using the loss of each network to improve the other. By effectively encoding the structural and semantic information from the code, these models can generate accurate and informative summaries and comments, facilitating code comprehension and maintainability.
        
\subsection{Probing in Natural Language Processing} \label{subsec:background:probing_in_nlp}
    Probing is used in NLP as a methodology for assessing the linguistic capabilities learned by ML models \cite{choudhary2022interpretation}. In general, probing involves evaluating how well the model embeddings capture syntactic \cite{hewitt2019structural}, semantic \cite{yaghoobzadeh2019probing}, or other types of linguistic properties inherent in language data \cite{pimentel2020information, goodwin2020probing}. Probes are classifiers trained to predict specific features based on learned representations generated by trained models and a high classification accuracy indicates that the target feature is sufficiently encoded in those representations \cite{vilone2020explainable}. NLP probes can be categorized into several types, each focusing on different linguistic aspects \cite{belinkov2022probing, luo2021local}:
    \begin{itemize}
        \item \textit{Syntactic probes}: These probes assess whether the model representations capture syntactic properties of language, such as part-of-speech tags, constituent or dependency parsing structures, and word order information.
        
        \item \textit{Semantic probes}: These probes aim to evaluate the model's ability to encode semantic information, such as semantic roles, and lexical relations.
        
        \item \textit{Coreference probes}: Coreference resolution is the task of identifying and linking mentions that refer to the same entity within a text \cite{mitkov2012coreference}. Probing for coreference capabilities helps understanding if the model can capture long-range dependencies and resolve ambiguities.
    \end{itemize}
    
    By systematically probing for these different linguistic properties, researchers can gain insights into the strengths and weaknesses of various model architectures and investigate the specific representations learned by the models, increasing interpretability and reliability \cite{belinkov2022probing}.

\subsection{AST-Probe} \label{subsec:background:ast_probe}
\begin{figure}[h]
        \centering
        \subfloat[\centering An example of a Python script \label{fig:ast_probe_snippet}]{{\includegraphics[width=0.25\textwidth,valign=c]{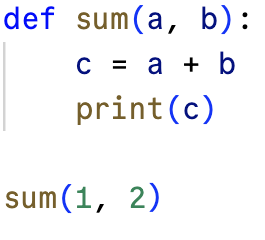} }}%
        \qquad
        \subfloat[\centering The AST constructed for the script \label{fig:astprobe_ast}]{{\includegraphics[width=0.65\textwidth,,valign=c]{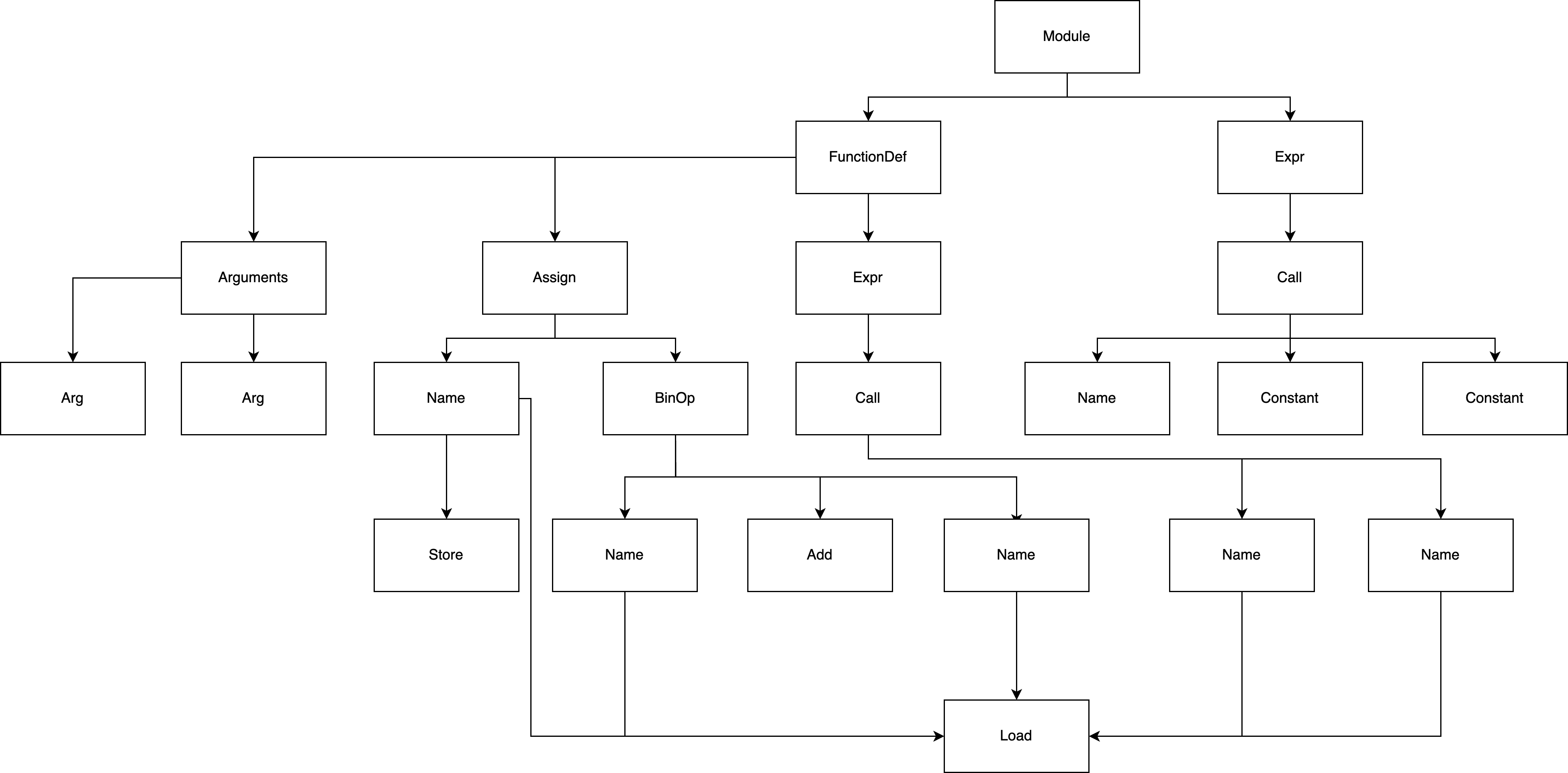} }}%
        \caption{An example of a Python script and its corresponding AST.}%
        \label{fig:motivating_example}%
    \end{figure}

    AST-Probe is designed to explore the latent space of LLMs and identify the subspace that encapsulates the syntactic information of programming languages \cite{hernandez2022ast}. 
    AST-Probe consists of the following steps: the AST of an input code snippet is constructed and 
    transformed into a binary tree. Next, the binary tree is represented as a tuple of \dcu vectors. It must be noted that the transformation from AST to binary tree, and from binary tree to \dcu is bi-directional, meaning that by having the \dcu tuple, one can reconstruct the AST of the input code. Since the AST constructed from a code snippet is syntactically valid (i.e., AST can only be constructed if code is syntactically correct), predicting the \dcu tuple from the representations extracted from the hidden layers of the model, given a code snippet as input, indicates that the model is capable of representing the syntax of the programming language in its latent space. 
    Figure \ref{fig:astprobe_ast} displays the AST extracted from a Python code snippet in Figure \ref{fig:ast_probe_snippet}. The \dcu tuple is constructed by parsing the binary tree. In this tuple, $d$ contains the structural information of the AST while $c$ and $u$ encode the labeling information. Algorithm \ref{algo:ast_probe_dcu_construction} describes the construction of the \dcu tuple in detail. 
        
    \begin{algorithm}
    \caption{Constructing the $<d, c, u>$ tuple for AST-Probe\cite{hernandez2022ast}}
    \begin{algorithmic}[1]
        \Procedure {TREE2TUPLE}{node}
            \If{node is leaf}
                \State $d \leftarrow []$
                \State $c \leftarrow []$
                \State $h \leftarrow 0$
                \If{node has unary label}
                    \State $u \leftarrow [\text{node.unary\_label}]$
                \Else
                    \State $u \leftarrow \theta$
            \EndIf
            \Else
                \State $l, r \leftarrow$ children of node
                \State $d_l, c_l, h_l \leftarrow$ TREE2TUPLE($l$)
                \State $d_r, c_r, h_r \leftarrow$ TREE2TUPLE($r$)
                \State $h \leftarrow \max(h_l, h_r) + 1$
                \State $d \leftarrow d + [h] + d_l + d_r$
                \State $c \leftarrow c + [\text{node.label}] + c_l + c_r$
                \State $u \leftarrow u + u_l + u_r$
            \EndIf
            \State \textbf{return} $d, c, u, h$
        \EndProcedure
    \end{algorithmic}
    \label{algo:ast_probe_dcu_construction}
    \end{algorithm}
    
    After constructing the \dcu of an input code snippet $I$, the embeddings of the model $M$ for the code are extracted from different layers. Suppose that model $M$ has $n$ layers: $l_{1}, ..., l_{n}$. We denote the output of the $l_{th}$ layer for input $I$ as $M(I)_{l}$ with $l$ being any layer between the second and the layer preceding the final layer (i.e., $2 < l < n-1$). After extracting $M(I)_{l}$ for an input code, the hidden representations are projected into a syntactic subspace $S$, and the probe is tasked with predicting the \dcu from this projection. This process is repeated for all layers and all the code samples in the training data. If, for a given dataset of codes, the \dcu tuples can be predicted by the probe with high accuracy, it indicates that 
    the model is capable of representing the syntax of the programming language in its latent space. 

\section{DeepCodeProbe}\label{sec:dcp} 
    In this section, we first present \dcp and how we represent AST/CFG of codes for probing the models. Afterward, we discuss our methodology for validating our proposed data representation and probing approach (i.e., DeepCodeProbe). 

\subsection{Feature Representation and Embedding Extraction} \label{subsec:dcp:proposed_approach}
    As described in Section \ref{subsec:background:probing_in_nlp}, probing is used in NLP to gain an understanding of what the model learns from its training data and analyze how the model makes a decision based on each input \cite{belinkov2022probing}. Furthermore, probing can also be used as an interpretability approach to validate whether the model learns representations from the data required for solving the task at hand \cite{choudhary2022interpretation}. Previous works \cite{hernandez2022ast, wan2022they, troshin2022probing, karmakar2021pre} have explored the ability of large language models (LLMs) to encode syntactic information of programming languages within their latent spaces, using probing.
     
    In this work, we are interested in DL models that are not as large as LLMs (i.e., language models trained on code that do not have millions/billions of parameters) and focus on smaller models that are trained on code for software maintenance tasks. Our proposed approach is probing such models on their capability of representing the syntax of the programming languages in their latent space. As such, we base our approach (DeepCodeProbe) on AST-Probe \cite{hernandez2022ast} as described in Section \ref{subsec:background:ast_probe}, to probe the models under study for syntactic information.

\begin{figure}
        \centering
        \includegraphics[width=0.7\linewidth]{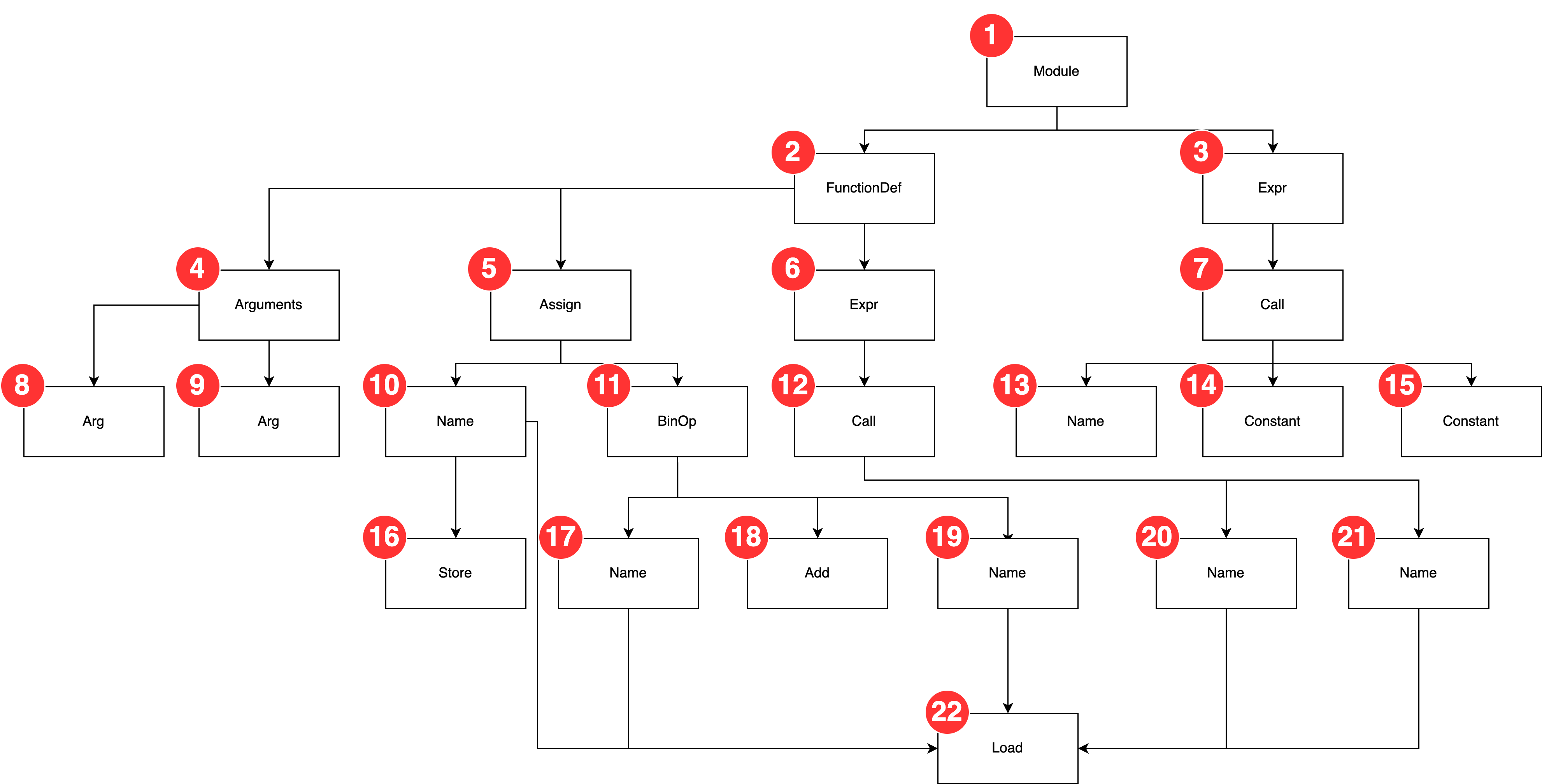}
        \caption{Annotated AST}
        \label{fig:dcp_annotated_ast}
    \end{figure}

    By basing our probing approach on the same principles of AST-Probe, we first extract a syntactically valid representation from the input code and construct a corresponding \dcu tuple from it. However, unlike AST-Probe, in our approach, we do not derive the \dcu tuple from the binary tree and instead build it directly from either the AST or the CFG extracted from the input code. We frame our probing approach as such since the binary tree of an AST or CFG would be very large and therefore, a 
    \dcu tuple constructed from it would be extremely high dimensional. Since we are interested in models that are much smaller than LLMs, probing these models for such high-dimensional representations will result in sparse representations of vectors (e.g., deriving a representation with 4,000+ features from an input with only 128 features). This sparsity would make the probe extremely 
    sensitive to noise \cite{lecun2015deep}. Given that the probe's capacity should be kept as small as possible (i.e. the probe should have significantly fewer parameters than the model being probed) \cite{hernandez2022ast, hewitt2019structural, belinkov2022probing}, such sparse representations of vectors will result in non-convergence because of the accumulation of small effects of noise in high dimensional data \cite{johnstone2009statistical, vandaele2022curse}. Therefore, we define the \dcu tuple construction from the AST or CFG of a given code as follows:

    \begin{itemize}
        \item $d$: The position of each node in the AST/CFG in sequential order level by level (breadth-wise) from left to right, as shown in Figure \ref{fig:dcp_annotated_ast}.
        \item $c$: Position of the children of each node in sequential order for ASTs and the connections between each node in CFG.
        \item $u$: Vector representation of each node in the AST/CFG extracted from the model's data processing approach.
    \end{itemize}
    
    Algorithm \ref{algo:scp_probing_overview} describes an overview of our probing approach. Similar to AST-Probe, extracting the \dcu tuple from an AST/CFG is bi-directional, meaning that the \dcu tuple extracted from the AST/CFG of a code can be used to reconstruct the corresponding AST/CFG. As this conversion is bi-directional, we infer that if the probe can predict the \dcu tuple from the embedding extracted from a model, then the model is capable of representing the syntax of the programming language in its latent space. In the same manner, we consider low performance on predicting the \dcu tuple, as the model's incapability to represent the syntax of the programming language in its latent space~\cite{hernandez2022ast, vilone2020explainable}.
    \newpage
    \begin{lstlisting}[caption=Generated \dcu tuple for the AST in Figure \ref{fig:dcp_annot
    ated_ast}, label={listing:dcu_example}]
        d = [ 1, 2, 3, 4, 5, 6, ....., 22 ]
        c = [ 
                (2, 3) # children of node 1.
                (4, 5, 6) # children of node 2,
                (7) # children of node 3,
                (8, 9) # children of node 4,
                .......
            ]
        u = [
                1 # Label of the ''Module'' node according to the model under study,
                54, # Label of the ''FunctionDef'' node according to the model under study,
                32, # Label of the ''Expr'' node according to the model under study,
                ......
            ]
    \end{lstlisting}
    
    \begin{algorithm}
        \caption{DeepCodeProbe}
        \begin{algorithmic}[1] 
        \small
            \Procedure{ExtractDCU}{$\text{SCRIPT}$} \Comment{Extract DCU tuple from code}
            \State \textbf{Input:} $\text{SCRIPT}$ \Comment{The input code script}
            \State \textbf{Output:} $\text{DCU}$ \Comment{The extracted DCU tuple}
            \If{$\text{MODEL\_INPUT\_TYPE} = \text{AST}$} \Comment{If model uses ASTs}
            \State $\text{SCRIPT\_AST} \gets \text{ConstructAST(SCRIPT)}$ \Comment{Construct AST}
            \ElsIf{$\text{MODEL\_INPUT\_TYPE} = \text{CFG}$} \Comment{If model uses CFGs}
            \State $\text{SCRIPT\_CFG} \gets \text{ConstructCFG(SCRIPT)}$ \Comment{Construct CFG}
            \EndIf
            \State $\text{SCRIPT\_DCU} \gets \text{Tree2Tuple(SCRIPT\_AST}$ \textbf{or} $\text{SCRIPT\_CFG)}$ \Comment{Convert AST/CFG to DCU tuple}
            \State \textbf{return} $\text{SCRIPT\_DCU}$
        \EndProcedure
        
        \Procedure{ProbeModel}{$\text{TRAINING\_DATA}$, $\text{MODEL}$} \Comment{Probe trained model}
            \State \textbf{Input:} $\text{TRAINING\_DATA}$ \Comment{Code used to train MODEL}
            \State \textbf{Input:} $\text{MODEL}$ \Comment{Trained model}
            \State \textbf{Output:} $\text{PROBING\_RESULTS}$ \Comment{Probing results}
            \State $\text{Probe} \gets \text{InitializeProbe()}$ \Comment{Initialize probe}
            \For{$\text{code} \in \text{TRAINING\_DATA}$}
                \State $\text{DCU} \gets \text{ExtractDCU(code)}$ \Comment{Extract DCU tuple}
                \State $\text{model\_embeddings} \gets \text{MODEL(code)}$ \Comment{Get model embeddings}
                \State $\text{Predicted\_DCU} \gets \text{Probe(model\_embeddings)}$ \Comment{Probe model}
                \State $\text{accuracy\_d}$, $\text{accuracy\_c}$, $\text{accuracy\_u} \gets \text{Compare(DCU, Predicted\_DCU)}$ \Comment{Compare predictions}
                \State $\text{PROBING\_RESULTS} \gets \text{PROBING\_RESULTS} \cup \{\text{accuracy\_d, accuracy\_c, accuracy\_u}\}$ \Comment{Store results}
            \EndFor
            \State \textbf{return} $\text{PROBING\_RESULTS}$
        \EndProcedure
        \end{algorithmic}
        \label{algo:scp_probing_overview}
    \end{algorithm}
    
    Figure \ref{fig:dcp_annotated_ast} displays the annotated AST that is used for \dcu tuple construction (the code snippet is shown in Figure \ref{fig:ast_probe_snippet} and the constructed AST in Figure \ref{fig:astprobe_ast}). The red circles above each node indicate the assigned index with the indexing being done level by level (breadth-wise) from left to right. Following this running example, Listing \ref{listing:dcu_example} displays the constructed \dcu tuple which will be used for probing the model.

\subsection{Validating \dcp} \label{subsec:dcp:validating_dcp}
    To ensure the reliability and accuracy of our probing task, we validate our proposed approach through the following three stages:
    
    \begin{itemize}
        \item Validation of the data representation method: We validate that the $<d, c, u>$ tuple constructed for each AST/CFG contains information on the programming language's syntax. 
        
        \item Validation of the extracted embeddings: We validate that the embeddings extracted from the model contain enough \textit{distinctive} information from the input. This is crucial to ensure that embeddings extracted from the model, for codes with dissimilar ASTs/CFGs are distinctly different from those extracted for codes with similar ASTs/CFGs.
        
        \item Validation that the embeddings contain information related to the task for which they are being probed: We validate that the models under study learn a relevant representation from their training data for the task for which they have been trained. This is crucial to ensure that the models under probe have not simply memorized their training data.
    \end{itemize}

     This validation is necessary because if the data representation approach or the extracted embeddings do not contain the information for which the model is being probed for, the results of the probing process will be meaningless since we would be probing the models for
     information that is not available in their latent space. Furthermore, if a model's embeddings do not change to reflect the task for which it has been trained (before and after training), these embeddings cannot be used for probing, since they would not contain enough information. In the following, we explain each of the aforementioned validation stages in more detail.

     \subsubsection{Validating data representation} \label{subsubsec:dcp:validating_dcp:validating_data_repr} DeepCodeProbe is designed for probing models that use syntactically valid constructs derived from code to accomplish their task. Previous works have shown that AST/CFG representations of code can be used to detect similarity between codes, as code clones of Types 1, 2, and weakly 3 have similar AST/CFG structures \cite{chilowicz2009syntax, tufano2018deep, svacina2020semantic, ain2019systematic}. Given this, we assume that the \dcu tuples for codes that are exact or near clones should be similar, while those for non-clones should be dissimilar. To validate this assumption, we construct \dcu tuples from the AST/CFG of code clones and non-clone pairs and measure the cosine similarity between these tuples. Similarity is defined as the relative distance between clone and non-clone pairs in the \dcu vector space. Since DeepCodeProbe works with vectors derived from tree and graph representations, using the cosine similarity metric allows us to assess the degree of similarity and dissimilarity between these vectors effectively. If the \dcu tuples for similar codes (clones of Type 1 and 2) are significantly more similar to each other compared to those for dissimilar codes (clones of Type 3 and 4), we can conclude that the data representation is valid.

     \subsubsection{Validating the extracted embeddings} \label{subsubsec:dcp:validating_dcp:validating_embeddings} Previous works have shown that semantic similarity between data points is reflected in embedding proximity in DL models (i.e. if two inputs are similar, then their representations in the model's latent space are similar) \cite{d2022spotlight, sohoni2020no}. Therefore, in order to validate the extracted embeddings we follow a similar approach to validating the data representation. In this step, instead of comparing the similarity of \dcu tuples, we compare the similarity of the extracted embeddings. We use the same code clone tuples constructed for validating the data representation. As such, we input code clone and non-clone pairs to each model and extract the corresponding embeddings for each input. Afterward, we use cosine similarity to measure the distance of the model's embeddings between code clones and non-clones. Given that the models under study are trained on syntactically valid representations of code, then regardless of the task that they are trained for, we expect that the extracted embeddings for code tuples that are syntactically similar (Types 1 and 2) to be similar to each other compared to code tuples that are not (Types 3 and 4). In other words, the cosine similarity for the embedding of similar codes should be significantly higher than the cosine similarity of dissimilar codes for the extracted embeddings to be valid. 

     \subsubsection{Validating the embedded information} \label{subsubsec:dcp:validating_dcp:validating_embedded_information}
     In probing literature, it is standard to show that the probe itself is incapable of learning representations from the extracted embeddings and instead only exposes information that already exists within the model's latent space \cite{hernandez2022ast, wan2022they, hewitt2019structural}. This is done by training the probe on embeddings extracted from randomly initialized versions of the model (i.e. models before training) and embeddings extracted from the model after it has been trained. If there exists a significant difference in the probe's performance in predicting the property of interest, then it is assumed that the probe is incapable of learning representations from the embeddings itself \cite{vilone2020explainable,belinkov2022probing, luo2021local,hernandez2022ast}. In our context, where we are probing small models trained on syntactic representations of code, we need to ensure that the probe's validity does not rely on the models' capacity to learn syntax. To address this, we propose an alternative validation method focusing on the similarity of embeddings for code clones, regardless of the models' syntax learning capacity. Given that the models are trained on syntactic representations of code, we extract embeddings for code clone tuples from both the trained and randomly initialized versions of the models. We then compare the cosine similarity scores for code clones from the trained model against those from the randomly initialized model. Higher similarity scores for the trained model would suggest that the model has learned some form of meaningful representations, even if not explicitly syntactic. Next, we train the probe on these embeddings and assess its performance. A significant difference in the probe's performance between embeddings from the trained and randomly initialized models would indicate that the probe is uncovering the existing structure in the embeddings rather than learning it during the probing process.
     \vspace{12pt}
     
     This comprehensive validation methodology ensures that our probe operates on reliable data representations and embeddings. By validating the data representation, we confirm that our \dcu tuples accurately capture the syntactic structures of codes. Through validating the extracted embeddings, we ensure that the models reflect meaningful semantic similarities. Finally, by validating the probe, we establish that it exposes the pre-existing information within the model's latent space without learning representations itself. These steps collectively validate the data representation and the probing approach, allowing us to analyze and draw conclusions about the models' capabilities in learning syntactic properties of code.

\section{Experimental Design}\label{sec:exp_design}
    In this Section, we will describe our experimental design. To conduct our study, we define the following research questions:

\begin{itemize}
    \item \textbf{RQ1}: Are models trained on syntactical representations of code capable of learning the syntax of the programming language?
    
    \item \textbf{RQ2}: In cases where models trained on syntactical representations of code fail to learn the programming language's syntax, do they learn abstract patterns based on syntax?
    
    \item \textbf{RQ3}: Does increasing the capacity of models without changing their architecture improve syntax learning capabilities?
    
\end{itemize}

In the rest of this section, we will first describe the models we have chosen for our study. Afterward, we describe the approaches used to construct the dataset for probing each model and lay out the details of the probes used for the models under study for answering RQ1 and RQ2. Finally, we describe how we increase the capacity of the models under study for answering RQ3.

\subsection{Models Under Study} \label{subsec:experiment_design:model_under_study}
    In this study, we are interested in models trained on code for software maintenance tasks. As such, to answer our RQs, we select two tasks in software maintenance for which learning the syntax of the programming language is considered to be important: Code Clone Detection (CCD) and code summarization. For each of these tasks, we select two models based on the following criteria:
    
    \begin{itemize}
        \item They are not large language models (LLMs).
        
        \item They are trained on syntactically valid representations of code, such as Abstract Syntax Trees (ASTs) or Control Flow Graphs (CFGs), rather than raw code or code artifacts as text.

        \item They demonstrate high performance on benchmarks specifically designed for the tasks they have been trained for.
    \end{itemize}
    
    Considering the criteria defined above, we select the following models to assess our probing approach, DeepCodeProbe:
    
    \subsubsection{AST-NN} \label{subsubsec:experiment_design:model_under_study:ast_nn} 
        Zhang et. al.\cite{zhang2019novel} proposed AST-NN as a novel approach for representing the AST extracted from code as inputs for DL models. In their approach, they break down the AST into Statement Trees (ST) to address the issues of token limitations and long-term dependencies in DL models, which arise from the large size of ASTs. This process involves a preorder traversal of the AST to identify and separate statement nodes defined by the programming language. For nested statements, statement headers and included statements are split to form individual statement trees. These trees, which may have multiple children (i.e., multi-way trees), represent the lexical and syntactical structure of individual statements, excluding sub-statements that belong to nested structures within the statement. This process reduces an AST into multiple smaller STs. Additionally, AST-NN employs a Word2Vec model \cite{mikolov2013distributed}, to encode the labeling information for each node in the AST by learning vector representations of words from their context in a corpus.
        
        \begin{figure}
            \centering
            \includegraphics[width=0.5\linewidth]{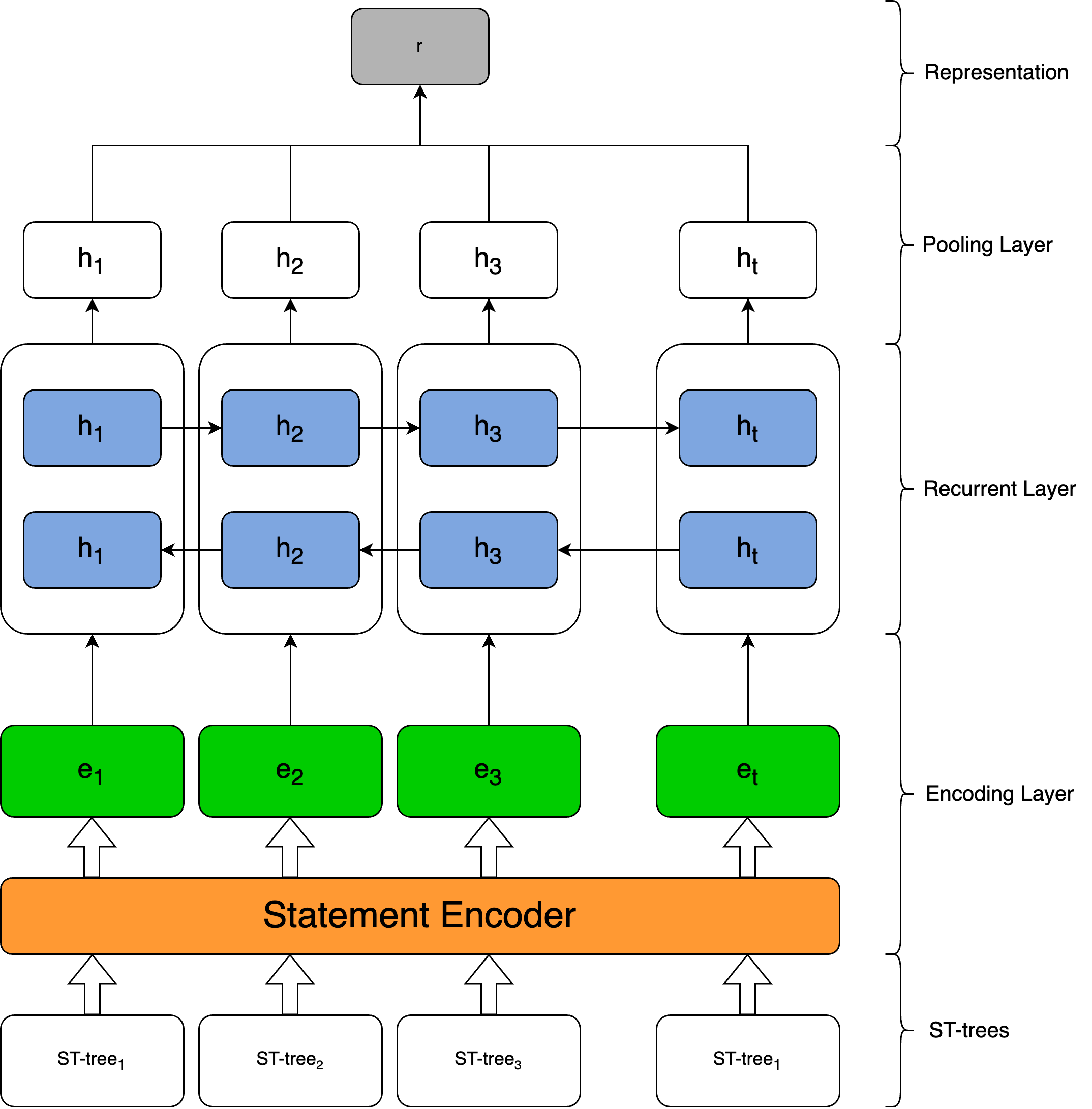}
            \caption{Model architecture for AST-NN \cite{zhang2019novel}}
            \label{fig:ed__astnn_arch}
        \end{figure}
        
        In AST-NN, the input code is broken down into STs to be used for two tasks, namely, source code classification and CCD. In this task, the STs extracted from two pieces of code are used as inputs for a DNN comprising an encoding layer, a recurrent layer, and a pooling layer. Figure \ref{fig:ed__astnn_arch} provides an overview of the AST-NN model architecture. The output of the representation layer is a binary value indicating whether two input codes are clones. The authors of AST-NN reported state-of-the-art clone detection results on the BigCloneBench~\cite{svajlenko2014towards} and OJClone \cite{mou2016convolutional} datasets for C and Java programming languages using their proposed approach.

    \subsubsection{FuncGNN} \label{subsubsec:experiment_design:model_under_study:func_gnn} 
\begin{figure}[h]
            \centering
            \includegraphics[width=0.8\linewidth]{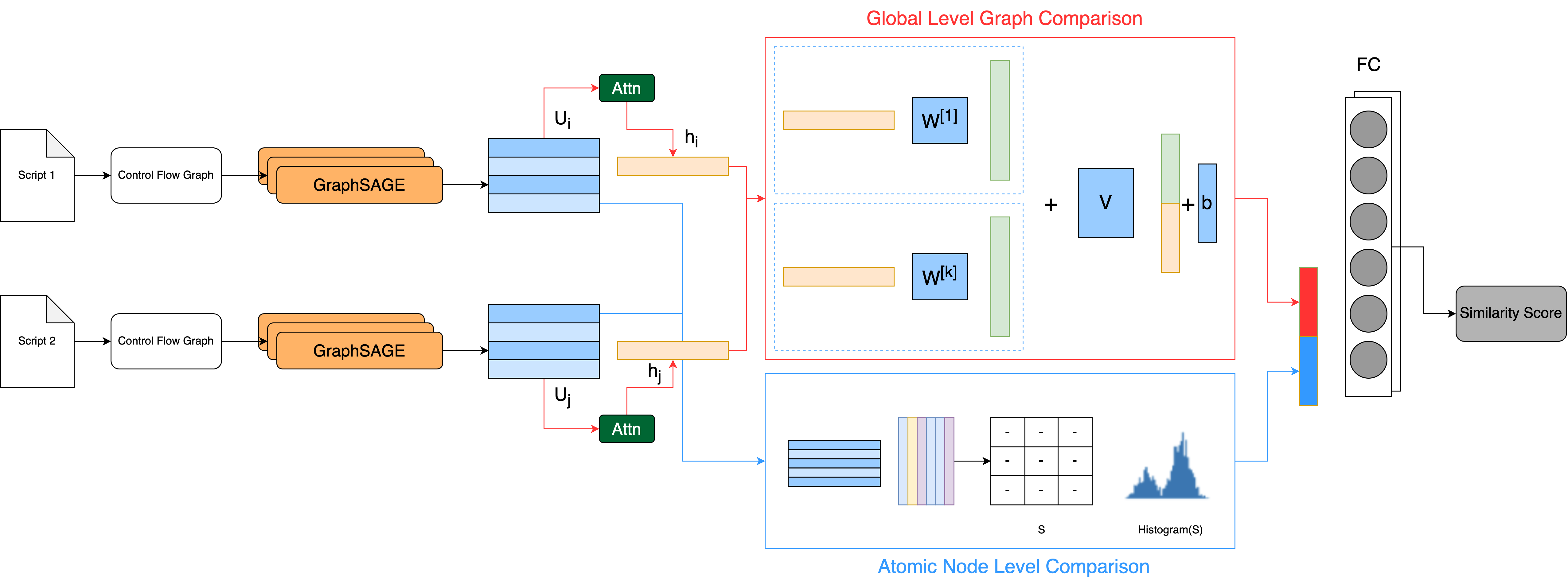}
            \caption{Model architecture for FuncGNN \cite{nair2020funcgnn}}
            \label{fig:ed__funcgnn_arch}
        \end{figure}

        Similar to AST-NN, Nair et. al. \cite{nair2020funcgnn} proposed FuncGNN, a graph neural network that predicts the Graph Edit Distance (GED) between the CFGs of two code snippets to identify clone pairs. FuncGNN, as presented in Figure \ref{fig:ed__funcgnn_arch}, operates by extracting CFGs from input code snippets and utilizing a statement-level tokenization approach, which distinguishes it from the word-level tokenization used by other models under study, such as AST-NN, SummarizationTF, and CodeSumDRL. The core assumption of FuncGNN is that code clone pairs will exhibit a lower GED compared to non-clone pairs, leveraging this metric to determine whether two code samples are clones.
     
        FuncGNN's methodology combines a top-down approach for embedding the global graph information with a bottom-up one for atomic-level node comparison to catch similarities between operations. The top-down approach generates an embedding that captures the overall structure and control flow of a CFG by incorporating an attention mechanism to find nodes in the CFG that have high semantic similarity. On the other hand, the bottom-up approach, focuses on the comparison of node-level similarities within the CFGs, to capture the similarities in program operations. By incorporating both approaches, FuncGNN predicts the GED between two input CFGs.

    \subsubsection{SummarizationTF} \label{subsubsec:experiment_design:model_under_study:sum_tf} 
\begin{figure}[h]
            \centering
            \includegraphics[width=0.8\linewidth]{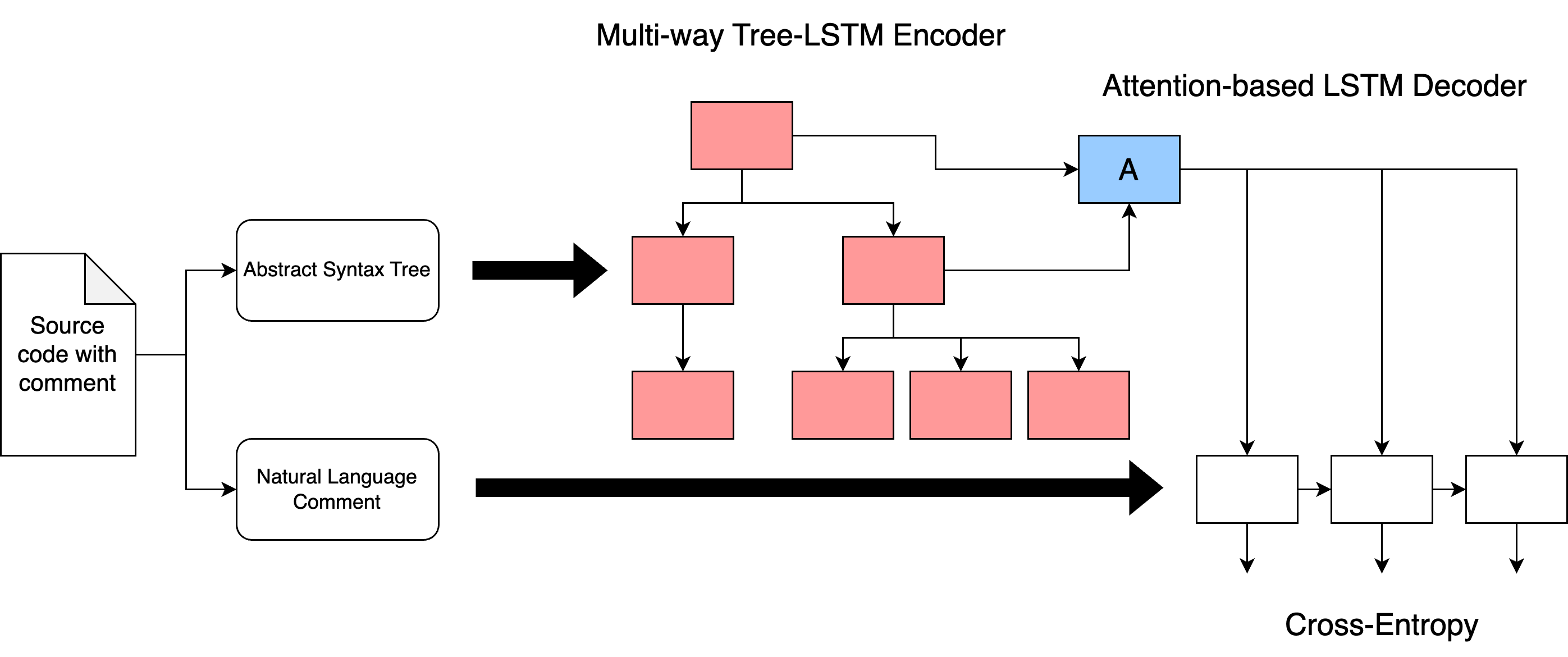}
            \caption{Model architecture for SummarizationTF \cite{shido2019automatic}}
            \label{fig:ed__sumtf_arch}
        \end{figure}

        The approach proposed by Shido et al.\cite{shido2019automatic} extends the 
        Tree-LSTM model \cite{tai2015improved} to handle source code summarization more effectively by introducing the Multi-way Tree-LSTM. This extension allows the model to process ASTs that have nodes with an arbitrary number of ordered children, which is a common feature in source code. In this approach, the AST of an input code is extracted and then each node in the AST is represented by a vector of fixed dimensions. These vectors are then used as input to the Multi-way Tree-LSTM and an LSTM decoder is then used to generate natural language summaries. Figure \ref{fig:ed__sumtf_arch} presents an overview of SummarizationTF's operation.

    \subsubsection{CodeSumDRL} \label{subsubsec:experiment_design:model_under_study:code_sum_drl} 
\begin{figure}
            \centering
            \includegraphics[width=0.8\linewidth]{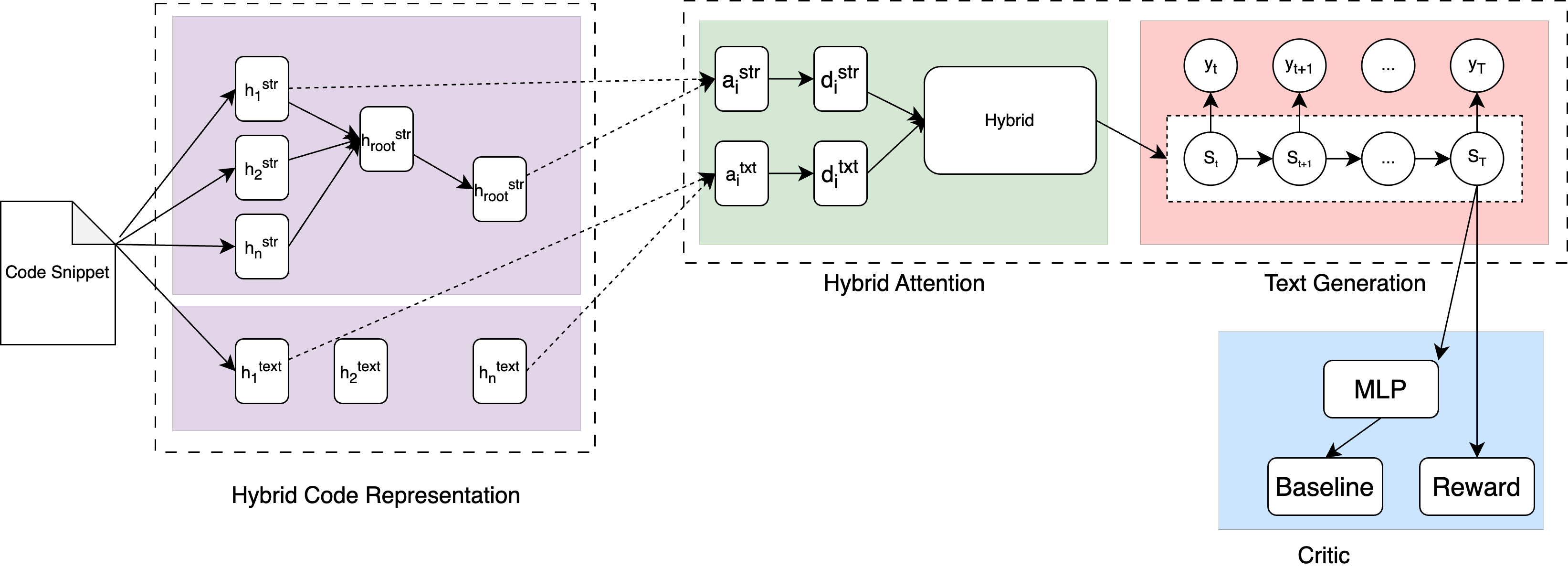}
            \caption{Model architecture for CodeSumDRL \cite{wan2018improving}}
            \label{fig:ed__codesumdrl_arch}
        \end{figure}

        Wan et. al.\cite{wan2018improving} proposed an approach for source code summarization based on Deep Reinforcement Learning (DRL). Their proposed approach integrates the AST structure and sequential content of code snippets as inputs to an actor-critic architecture as presented in Figure \ref{fig:ed__codesumdrl_arch}. The actor-network suggests the next best word to summarize the code based on the current state, providing local guidance, while the critic-network assesses the possibility of occurrence of all possible next states. Initially, both networks are pre-trained using supervised learning (i.e., code and documentation tuples), using cross-entropy and mean square as loss functions, for the actor and the critic respectively. Afterward, both networks are further trained through policy gradient methods using the BLEU metric as the advantage reward to improve the model's performance.
        To train the model, code is represented in a hybrid manner in two levels: lexical level (code comments) and syntactic level (code ASTs) \cite{wan2018improving}. An LSTM is used to convert the natural language comments and a separate Tree-based-RNN is used to represent the code ASTs. The output of these two layers is then used to represent the input code alongside its comment for training the actor and critic.
        
    For each of the models described above we use the code and data in the replication packages provided by the authors and train the models from scratch to replicate the results of the original studies \cite{zhang2019novel, nair2020funcgnn, shido2019automatic, wan2018improving}. It should be noted that for each model, we do not modify the codes or the data in the replication packages and in cases where it was required (e.g., deprecated libraries, syntax errors in the code, etc.) we kept our modifications as minimal as possible. We provide the source code of each model alongside our probing approach in our replication package \cite{rep_package}.

\subsection{Dataset Construction for Each Model} \label{subsec:experiment_design:dataset_construction}
    After training models and replicating the results of the original studies, we use the same data that was used in the original works as input to the models and at the same time extract the AST/CFG from each input and construct the $<d, c, u,>$ tuple which was explained in Section \ref{subsec:dcp:proposed_approach}. Afterward, we extract the intermediate outputs of the model for the given input and train the probe to predict the constructed $<d, c, u,>$ tuple from the extracted embeddings. It is important to note that the chosen models contain multiple hidden layers. Therefore, for each model, we extract the embeddings from the layers that provide the most information-rich representations learned by the model, as follows:
    
    \begin{itemize}
        \item AST-NN: The AST-NN model architecture, as shown in Figure \ref{fig:ed__astnn_arch}, consists of three main components: an encoding layer, a recurrent layer, and a pooling layer. The encoding layer first converts the input STs as described in Section \ref{subsubsec:experiment_design:model_under_study:ast_nn} into numerical representations suitable for the model. These encoded inputs are then processed by the recurrent layer. The outputs of the recurrent layer are subsequently passed through the pooling layer, which aggregates the representations and produces a fixed-size vector encoding of? the entire input STs. For probing, we extract the embeddings from both the recurrent and the pooling layers.
        
        \item FuncGNN: The FuncGNN model architecture, as illustrated in Figure \ref{fig:ed__funcgnn_arch}, consists of several components: a graph convolution layer, an attention mask mechanism, and a ReLU activation function. The graph convolution layer learns representations by capturing the structural information from the input CFGs. These learned representations are then passed through the attention mask, which assigns importance weights to different parts of the outputs of the graph convolution layer. Finally, a ReLU activation function is applied to the attention-weighted representations, and a sigmoid activation yields a binary output indicating whether the two input CFGs are code clones or not. For probing, we extract the embeddings from the output of the graph convolution layer.
        
        \item SummarizationTF: As displayed in Figure \ref{fig:ed__sumtf_arch}, the SummarizationTF model follows an encoder-decoder architecture. The encoder component processes the AST generated from a code snippet and generates a contextualized representation from it. The outputs of the encoder layer are used as input to the decoder layer to generate summaries from the input AST. For probing, we extract the embeddings from the outputs of the encoder component.
        
        \item CodeSumDRL: The CodeSumDRL model employs a dual-encoder architecture as displayed in Figure \ref{fig:ed__codesumdrl_arch}, where separate LSTM layers are used to encode the input code snippet and the associated comments. The outputs of these LSTM layers are then passed as inputs to an encoder layer. For probing, we extract the embeddings from the outputs of the encoder layer.
    \end{itemize}
    
    The rationale behind selecting these specific layers is that they are expected to capture the most informative and high-level representations of the input data, as they are the final representations before subsequent transformations or output generations. After extracting the embeddings, the dataset used for training \dcp for each model consists of (\texttt{embeddings}, \dcu) tuples from the inputs given to each model. We outline the detailed dataset construction for each of the models under study as follows.

    \subsubsection{AST-NN} \label{subsubsec:experiment_design:dataset_construction:ast_nn}
        Algorithm \ref{algo:ed__astnn_code_to_repr} presents how we construct the \dcu tuples for probing AST-NN using DeepCodeProbe. Note that to represent the labels of each node, we use the same Word2Vec model that is used by AST-NN to convert the labels into numerical representations (i.e., indexes) as described in Section \ref{subsubsec:experiment_design:model_under_study:ast_nn}. We probe this model on its capability to represent the syntax of both C and Java programming languages. We also adopt the same preprocessing approach as AST-NN to convert the inputs from code to STs. Afterward, for each ST, we construct the \dcu tuple to be used for training the probe.

        \begin{algorithm}
            \caption{Generating \dcu Tuples from STs}
            \begin{algorithmic}[1]
            \Require An Abstract Syntax Tree (AST) $tree$
            \Ensure Dictionary containing depth (D), child indices (C), and unary label (U) tuples for ST blocks
            
            \Procedure{Tree2Tuple}{$tree$}
                \State $word2VecEmbeddings \leftarrow$ Load appropriate Word2Vec model based on language
                \State $maxTokenIndex \leftarrow$ Number of vectors in $word2VecEmbeddings$
                \State $vocab \leftarrow word2VecEmbeddings$
                
                \Function{ConvertNodeToIndex}{$node$}
                    \State $tokenIndex \leftarrow$ Index of $node.token$ in $vocab$ or $maxTokenIndex$ if not found
                    \State $childIndices \leftarrow$ Empty list
                    \For{$child$ in $node.children$}
                        \State $childIndices$.append(\Call{ConvertNodeToIndex}{$child$})
                    \EndFor
                    \State \Return $\left[ tokenIndex \right] + childIndices$
                \EndFunction
                
                \Function{Tree2Index}{$rootNode$}
                    \State $blocks \leftarrow$ Extract ST blocks from AST starting from $rootNode$
                    \State $sequenceInfo \leftarrow$ Initialize empty dictionary for block info
                    \For{$i = 0$ to $length(blocks) - 1$}
                        \State $block \leftarrow blocks[i]$
                        \State $blockTokenIndex \leftarrow$ Index of $block.token$ in $vocab$ or $maxTokenIndex$
                        \State $blockChildrenIndices \leftarrow$ \Call{ConvertNodeToIndex}{$block$}
                        \State $sequenceInfo[i].d \leftarrow i + 1$
                        \State $sequenceInfo[i].c \leftarrow$ Flatten $blockChildrenIndices$ excluding the first element
                        \State $sequenceInfo[i].u \leftarrow blockTokenIndex$
                    \EndFor
                    \State $D \leftarrow$ Extract $d$ values from $sequenceInfo$ for all blocks
                    \State $C \leftarrow$ Extract $c$ values from $sequenceInfo$ for all blocks
                    \State $U \leftarrow$ Extract $u$ values from $sequenceInfo$ for all blocks
                    \State \Return $\{ d: D, c: C, u: U \}$
                \EndFunction
            
                \State \Return \Call{Tree2Index}{$tree.root$}
            \EndProcedure
            \end{algorithmic}
            \label{algo:ed__astnn_code_to_repr}
        \end{algorithm}
        
    \subsubsection{FuncGNN} \label{subsubsec:experiment_design:dataset_construction:func_gnn}
        Algorithm \ref{algo:ed__funcgnn_code_to_repr} presents an overview of the \dcu tuple construction for FuncGNN in DeepCodeProbe. The training data for FuncGNN is already pre-processed from code into CFGs using Soot\footnote{https://github.com/Sable/soot}, a framework that is used to transform Java bytecode into intermediate representation for analysis, visualization, and optimization. In this format, each node represents a line in the program, and the edges between the nodes indicate the control flow between each node. For constructing the \dcu tuple, we traverse the CFG and use the depth, connections, and labels of each node to represent it in a numerical vector for training and evaluating the probe. 
       
        \begin{algorithm}
            \caption{Generating \dcu Tuples from CFGs}
            \begin{algorithmic}[1]
            \Require Code graphs represented as lists of edges and node labels
            \Ensure Tuples representing depth (D), connections (C), and unary label (U) for the code graphs
            \Function{CFG2Tuple}{$codeGraphs, nodeLabelMappings$}
                \For{$codeGraph$ in $codeGraphs$}
                    \State $edges \gets codeGraph.edges$
                    \State $reverseEdges \gets $ list of edges with source and target swapped
                    \State $edgeList \gets edges + reverseEdges$ \Comment{Ensure connections in both directions}
                    \State Convert $edgeList$ to an appropriate data structure
                    \State $depths \gets $ list of node depths in the graph traversal order
                    \State $connections \gets edgeList$
                    \State $usages \gets $ list of node label mappings from $nodeLabelMappings$
                    \State Store $(depths, connections, usages)$ as a tuple for the current $codeGraph$
                \EndFor
                \State \Return list of $(depths, connections, usages)$ tuples for all code graphs
            \EndFunction
            \end{algorithmic}
            \label{algo:ed__funcgnn_code_to_repr}
        \end{algorithm}

    \subsubsection{SummarizationTF} \label{subsubsec:experiment_design:dataset_construction:sum_tf}
        This model works with ASTs. Therefore, unlike AST-NN there is no need to first generate the ASTs from code and then break them down into STs. As such, the process for generating the \dcu tuples is similar to that of Algorithm \ref{algo:ed__astnn_code_to_repr} with some modifications to the method of parsing the trees. Algorithm \ref{algo:ed__sumtf_code_to_repr} presents our approach in detail. 
        
        \begin{algorithm}
            \caption{Generating \dcu tuples from ASTs}
            \begin{algorithmic}[1]
            \Require Source code $code$
            \Ensure Tuples of node positions (D), children count (C), and unary label (U) for AST processing
        
            \Procedure{Tree2Tuple}{$AST$}
                \State $token2Index \leftarrow$ Load appropriate tokenizer based on the programming language
                \State $maxTokenIndex \leftarrow$ Highest index in $token2Index$
                \State $vocab \leftarrow token2Index$
        
                \Function{ConvertNodeToIndex}{$node$}
                    \State $tokenIndex \leftarrow$ Index of $node.token$ in $vocab$ or $maxTokenIndex$ if not found
                    \State $childIndices \leftarrow$ Empty list
                    \For{$child$ in $node.children$}
                        \State $childIndices$.append(\Call{ConvertNodeToIndex}{$child$})
                    \EndFor
                    \State \Return $\left[ tokenIndex \right] + childIndices$
                \EndFunction
        
                \Function{AST2Index}{$tree$}
                    \State $nodes \leftarrow$ Extract nodes from $tree$
                    \State $sequenceInfo \leftarrow$ Initialize empty dictionary for node info
                    \For{$i = 0$ to $length(nodes) - 1$}
                        \State $node \leftarrow nodes[i]$
                        \State $nodeTokenIndex \leftarrow$ Index of $node.type$ in $vocab$ or $maxTokenIndex$
                        \State $nodeChildrenIndices \leftarrow$ \Call{ConvertNodeToIndex}{$node$}
                        \State $sequenceInfo[i].d \leftarrow$ Extract positions for $node$ in AST
                        \State $sequenceInfo[i].c \leftarrow$ $nodeChildrenIndices$
                        \State $sequenceInfo[i].u \leftarrow nodeTokenIndex$
                    \EndFor
                    \State $D \leftarrow$ Extract position values from $sequenceInfo$ for all nodes
                    \State $C \leftarrow$ Extract children count from $sequenceInfo$ for all nodes
                    \State $U \leftarrow$ Extract type indices from $sequenceInfo$ for all nodes
                    \State \Return $\{ d: D, c: C, u: U \}$
                \EndFunction
        
                \State \Return \Call{AST2Index}{$AST$}
            \EndProcedure
            \end{algorithmic}
            \label{algo:ed__sumtf_code_to_repr}
        \end{algorithm}

    \subsubsection{CodeSumDRL} \label{subsubsec:experiment_design:dataset_construction:code_sum_drl}
        Similar to SummarizationTF, CodeSumDRL also works with ASTs generated from code. As such, we follow the same steps as outlined in Algorithm \ref{algo:ed__sumtf_code_to_repr} to build the necessary \dcu tuples for training and evaluating the probe.

\subsection{Probe's Design for Each Model} \label{subsec:experiment_design:probe_design_for_models}
    As explained in Section \ref{subsec:background:probing_in_nlp}, the probe should have minimal size and complexity. Following the same conventions used in probing language models \cite{maudslay2020tale}, the probe should not learn any information from the data but instead, learn a transformation (i.e., projection) of the embeddings that exposes the information that the model learns from its training data. Therefore, for each of the models under study, based on their capacity (i.e., number of parameters), we experiment with probes with different sizes. To prevent the probes from learning representations from the model's embeddings, we follow the same design approach as used in \cite{hernandez2022ast}. This means that each probe has a single hidden layer, and the number of units in this layer is adjusted based on the size and architecture of the model being probed. We include the code for training probes of different sizes for each of the models under study in our replication package \cite{rep_package}.

\subsection{Scaling Up The Models}\label{subsec:experiment_design:scaling_up_the_models}
    Kaplan et al. \cite{kaplan2020scaling} examined the scaling laws for language models, focusing on the impact of increasing model capacity, training data, and computational resources on the models' training efficiency and performance. In their study, they experiment with how increasing each and a combination of these factors affects the model's sample efficiency, overfitting, convergence, and overall performance on the test benchmarks. Their research focuses on transformer-based language models and is therefore not directly applicable to our work. However, we can apply the same principles to the models under study and investigate how increasing the models' capacity affects their syntax learning capabilities. Based on their work, to address RQ3, we keep the amount of compute and the dataset size the same, while increasing the number of models' parameters (i.e., capacity) without changing the models' architecture. We have described the architecture of each of the models under study alongside the layers from which we have extracted the embeddings in Sections \ref{subsec:experiment_design:model_under_study} and \ref{subsec:experiment_design:dataset_construction}, respectively. To align our approach with the principles outlined by Kaplan et al. \cite{kaplan2020scaling}, we define ``scaling up'' for each model in our study as increasing model capacity by expanding the number of units within the layers responsible for generating outputs used for probing. Specifically, our scaling strategy includes:
    
    \begin{itemize}
        \item AST-NN: Increasing the recurrent and pooling layers' unit count and leaving the encoding layer unchanged.
    
        \item FuncGNN: We increase the number of units of the graph convolution layer without altering the number of attention heads.
        
        \item SummarizationTF: The Multi-way Tree-LSTM's unit count is increased, while all other components remain the same.
        
        \item CodeSumDRL: The encoder's unit count is increased, with no modifications to the rest of the components.
    \end{itemize}
    
    This approach to scaling allows for a controlled alteration of the models' capacity, which in turn allows for a more precise study into how such modifications influence their syntax learning capabilities.

\section{Results and Analysis}\label{sec:results}
    Before analyzing the embeddings extracted from the models to assess their capability to learn the syntax of the programming language and answer our research question, we must first ensure the reliability of the probing approach used to obtain those embeddings. Therefore, in the following, we first conduct a validation of our probing approach before proceeding to answer our research question. 
\subsection{Evaluation of our Probing Approach \dcp} \label{subsubsec:experiment_results:rq1:validation}

 As described in Section \ref{subsec:dcp:validating_dcp}, we evaluate \dcp~ by first assessing the validity of the \dcu tuple and the embeddings extracted from the models. As AST-NN and FuncGNN are designed specifically for CCD, we use the same dataset provided by their authors (AST-NN \cite{zhang2019novel} and FuncGNN \cite{nair2020funcgnn}) for the validation of \dcp~ on these models. In the case of SummarizationTF and CodeSumDRL, the authors did not originally include a code clone dataset. Hence, to assess the validity of our probing approach on these models, we use the code clone dataset of AST-NN for SummarizationTF (as they are both trained on Java) and the Python code clone detection dataset from \cite{poolcpythondataset} which was used to train a code clone detector based on CodeBERT \cite{feng2020codebert} for Python code clone detection \cite{codebertpython}, for CodeSumDRL.
        
Table \ref{table:experiment_results::dcu_sim} presents the average cosine similarity for the \dcu tuples for each of the models under study. The ``Programming Language'' column specifies the programming language on which the model is trained. The ``Comparison criterion'' denotes which subsets of the datasets were used for the similarity comparison, with ``similar'' indicating comparison among code clones (Types 1 and 2) and ``dissimilar'' indicating comparison among non-clone pairs. Finally, the ``Average Similarity'' columns show the average cosine similarity across all \dcu tuples constructed for each model and comparison criterion. 
        
        \begin{table}[h]
            \caption{Results of Cosine similarity of \dcu tuples for each model}
            \centering
            \resizebox{\textwidth}{!}{
                \begin{tabular}{c c c r r r}
                    \toprule
                    Model & Programming Language & Similarity Criterion & Average Similarity - D(\%) & Average Similarity - C(\%) & Average Similarity - U(\%)\\
                    \midrule
                        \multirow{4}{*}{AST-NN} 
                        & \multirow{2}{*}{C} & Similar & 64 & 17 & 43 \\
                        & & Dissimilar & 57 & 12 & 4\\
                        \cline{3-6}
                        & \multirow{2}{*}{Java} & Similar & 80 & 50 & 76\\
                        & & Dissimilar & 42 & 16 & 54\\
                    \midrule
                        \multirow{2}{*}{FuncGNN}  
                        & \multirow{2}{*}{Java} & Similar & 90 & 86 & 82\\
                        & & Dissimilar & 57 & 36 & 68\\
                    \midrule
                        \multirow{2}{*}{SummarizationTF}  
                        & \multirow{2}{*}{Java} & Similar & 75 & 83 & 74\\
                        & & Dissimilar & 33 & 60 & 51\\
                    \midrule
                        \multirow{2}{*}{CodeSumDRL}  
                        & \multirow{2}{*}{Python} & Similar & 0 & 28 & 35\\
                        & & Dissimilar & -11 & 3 & 20\\
                    \bottomrule
                \end{tabular}
            }
            \label{table:experiment_results::dcu_sim}
        \end{table}
        
        As shown by the results in Table \ref{table:experiment_results::dcu_sim}, we can observe a significant difference in the \dcu similarities between clone and non-clone code pairs across all models. Specifically, for AST-NN we can observe a 7\% difference in the representations of $d$ tuples which encode node location, a 5\% difference in the representation of $c$ tuples which encode the node's children, and a major difference of 39\% between $u$ tuples which encode node labeling information which lines up with how large ASTs are broken down into smaller STs for this model \cite{zhang2019novel}. For FuncGNN, we observe noticeable differences in representations constructed for CFGs, showing a strong differentiation based on node position, connections, and labels between clone and non-clone pairs especially across the $d$ tuples which encode the position of each node in the CFG and the $c$ tuples which encode the connections between the nodes. We observe a similar trend for SummarizationTF, especially with a difference of 43\% for the $d$ tuple, 23\% for the $c$ tuple, and 23\% for the $u$ tuple which lines up with how ASTs of dissimilar codes have different structures as also described in Section \ref{subsubsec:dcp:validating_dcp:validating_data_repr}. For CodeSumDRL, we observe similar results as SummarizationTF in the similarities between the \dcu tuples as well as a similar difference of distinction between the encoded representation of the nodes' positions and children. Based on these results, we can infer that our \dcu tuple representation, extracted from ASTs/CFGs, holds sufficient syntactical information and can be used for probing for syntactic information on the models under study.

        With the validity of the constructed \dcu tuples established, we proceed to assess the validity of the information within the embeddings extracted from each model. Table \ref{table:experiment_results::embeddings_sim} presents the results of embedding validation. Following a similar methodology to validating the \dcu tuples representation, we compute the average cosine similarity across both clone and non-clone pairs. Additionally, the ``Trained'' column indicates whether the embeddings were extracted from an untrained or trained instance of the model.

        \begin{table}[h]
            \caption{Results of Cosine Similarity of embeddings before and after training}
            \centering
            \resizebox{0.85\textwidth}{!}{
            \begin{tabular}{c c c c r}
                \toprule
                Model & Programming Language & Similarity Criterion & Trained & Average Cosine Sim (\%)\\
                \midrule
                    \multirow{8}{*}{AST-NN} 
                    & \multirow{4}{*}{C} & \multirow{2}{*}{Similar} & Y & 28.89\\
                    & & & N & 4.81\\
                    \cline{3-5}
                    & & \multirow{2}{*}{Dissimilar} & Y & 20.38\\
                    & & & N & 1.90\\
                    \cline{2-5}
                    & \multirow{4}{*}{Java} & \multirow{2}{*}{Similar} & Y & 36.43\\
                    & & & N & 0.42\\
                    \cline{3-5}
                    & & \multirow{2}{*}{Dissimilar} & Y & 20.12\\
                    & & & N & 0.51\\
                \midrule
                    \multirow{4}{*}{FuncGNN}  
                    & \multirow{4}{*}{Java} & \multirow{2}{*}{Similar} & Y & 86.15\\
                    & & & N & 78.46\\
                    \cline{3-5}
                    & & \multirow{2}{*}{Dissimilar} & Y & 72.55\\
                    & & & N & 71.05\\
                \midrule
                    \multirow{4}{*}{SummarizationTF} 
                    & \multirow{4}{*}{Java} & \multirow{2}{*}{Similar} & Y & 59.33\\
                    & & & N & 52.90\\
                    \cline{3-5}
                    & & \multirow{2}{*}{Dissimilar } & Y & 53.47\\
                    & & & N & 43.97\\
                \midrule
                    \multirow{4}{*}{CodeSumDRL}  
                    & \multirow{4}{*}{Python} & \multirow{2}{*}{Similar} & Y & 19.03\\
                    & & & N & 8.71\\
                    \cline{3-5}
                    & & \multirow{2}{*}{Dissimilar} & Y & 11.60\\
                    & & & N & 3.19\\
                \bottomrule
            \end{tabular}
            }
            \label{table:experiment_results::embeddings_sim}
        \end{table}

        The analysis of data from Table \ref{table:experiment_results::embeddings_sim} reveals a significant distinction between the embeddings of clone and non-clone codes. Specifically, we observe that after training the models, the similarity of the extracted embeddings for code clones increases. Furthermore, the distance in the embeddings extracted from the models between clone and non-clone codes changes significantly post-training, as opposed to their randomly initialized stage. In more detail, for code clones, AST-NN shows a significant increase from 4.81\% to 28.89\% in C, and from 0.42\% to 36.43\% in Java, post-training. We observe an increase in the cosine similarity of non-clone codes as well, but to a lesser degree, which can be attributed to the model's ability to identify some form of structure in the STs which we expand more on, in RQ2 (Section \ref{subsec:experiment_results:rq2}). We observe a similar pattern for FuncGNN with the similarity of code clones increasing from 78.46\% to 86.15\%, and just a 1.5\% increase between the embedding similarity of non-clone codes, post-training. The marginal increase for non-clone codes shows the model's capability to recognize underlying structural patterns which we will expand on when discussing the probing results in Section \ref{subsec:experiment_results:rq1}. For SummarizationTF, we also observe the same pattern of average cosine similarity increasing for code clones after training the model with a lower cosine similarity for non-clone pairs. We observe a 6.4\% difference between the embeddings of code clones before and after training the model and a 9.5\% increase for non-clone codes. Finally, for CodeSumDRL we observe that the embedding similarity increases from 8.71\% to 19.03\% for code clones, and from 3.19\% to 11.6\% for non-clones. These results confirm the validity of the extracted embeddings from each model for probing, as they show that each model learns some information from the syntactic representations and does not memorize its training data.

        By cross-analysing the data in Tables \ref{table:experiment_results::dcu_sim} and \ref{table:experiment_results::embeddings_sim}, we can infer the following:
        \begin{itemize}
            \item The \dcu tuple, which is constructed for each model following the model's data representation approach, encodes the syntactic information in the AST/CFG. It should be noted that the \dcu tuple construction and its syntactic representation are separate from the model embeddings and the representations that the model learns from its training data.
            \item The models demonstrate an ability to learn some form of representation from the input AST/CFGs, enabling the differentiation between clone and non-clone code pairs even if they are not trained for CCD.
        \end{itemize}

\subsection{RQ1: Are models trained on syntactical representations of code capable of learning the syntax of the programming language?} \label{subsec:experiment_results:rq1}

    Table \ref{table:experiment_results::probe_results} presents the results of our probing as described in Sections \ref{subsec:experiment_design:dataset_construction} and \ref{subsec:experiment_design:probe_design_for_models}. The ``Accuracy-D'', ``Accuracy-C'', and ``Accuracy-U'' columns indicate the probe's accuracy in predicting the $d$, $c$, and $u$ tuples, respectively. As we can observe from the results, the models under study are incapable of representing the syntax of the programming language in their latent space. For the AST-NN model trained for CCD on C and Java, the poor accuracy scores of 8\% across all syntax tuples ($d$, $c$, and $u$) indicate an extremely limited ability to encode programming language's syntax in its latent space. Similarly, the SummarizationTF model for code summarization struggles with syntax learning as well, achieving low accuracies of 13.83\%, 14.79\%, and 14.79\% for the $d$, $c$, and $u$ tuples respectively. In contrast, the CodeSumDRL model exhibits a relatively stronger syntax learning ability, with accuracies of 41.6\% for $d$, 33.92\% for $c$, and 28.08\% for $u$ tuples on the code summarization task. While not close to the results reported for LLMs in \cite{hernandez2022ast, troshin2022probing, wan2022they} (where the probes exhibit over 80\% accuracy in retrieving information related to the programming language's syntax), these higher scores suggest that CodeSumDRL learns some syntactic patterns from code. The most intriguing result comes from FuncGNN's probe. As described in Section \ref{subsubsec:experiment_design:model_under_study:func_gnn}, FuncGNN was designed for CCD on CFGs. Remarkably, it achieves 98\% accuracy in predicting the $c$ tuple that encodes edge connection information in the code's CFG. Thus, we can conclude that FuncGNN particularly focuses on the edges of the CFGs (connections between nodes) in order to detect clones between two input codes. This observation also aligns
    with the results reported in \cite{nair2020funcgnn}, in which the goal of FuncGNN is to predict the GED between the CFGs of two input codes. However, its performance of 43.36\% and 39.26\% for $d$ and $u$ tuples indicate that the model is incapable of representing the full syntax of the programming language.
        
    \begin{table}[h]
        \caption{Result of DeepCodeProbe's accuracy on recovering Syntactic Information}
        \resizebox{\textwidth}{!}{
            \centering
            \begin{tabular}{c c c r r r}
                \toprule
                Model & Task & Programming Language & Accuracy-D(\%) & Accuracy-C(\%) & Accuracy-U(\%)\\
                \midrule
                    \multirow{2}{*}{AST-NN} & \multirow{2}{*}{CCD} & C & 8.65 & 8.63 & 8.65 \\
                     & & Java & 8.33 & 8.09 & 8.65\\
                \midrule
                    FuncGNN & CCD & Java & 43.36 & 98.51 & 39.26\\
                \midrule
                    SummarizationTF & Code Summarization & Java & 13.83 & 14.79 & 14.79\\
                \midrule
                    CodeSumDRL & Code Summarization & Python & 41.60 & 33.92 & 28.08\\
                \bottomrule
            \end{tabular}}
        
        \label{table:experiment_results::probe_results}
    \end{table}
    
    From the results of our probing, we can see that our probes are not capable of predicting the \dcu tuples for their input codes for AST-NN and SummarizationTF which as described in Section \ref{subsec:dcp:proposed_approach}, indicates that they are incapable of learning the syntax of the programming language that they are trained on. Even though in comparison to the mentioned models, CodeSumDRL's probe achieves higher accuracies across the \dcu tuples, the results show that its syntax learning capability is limited. However, as these model show good performance on the benchmarks that they were presented with and our validation shows that they do not memorize their training data, we need to investigate what representation they learn from code for achieving their task. We investigate this in RQ2 (Section \ref{subsec:experiment_results:rq2}).

    \begin{tcolorbox}[colback=blue!5,colframe=blue!40!black]
        \textbf{Findings 1:} By analyzing the embeddings extracted from the models under study for clone and non-clone pairs, we can observe that the models learn some information about the programming language's syntax. However, our probe shows that none of the models under study are capable of representing the complete syntax in their latent space. This indicates that the models under study are either incapable of representing the full syntax of the programming language in their latent space due to their limited capacity (as opposed to LLMs) or do not require the full syntax to achieve their objectives.
    \end{tcolorbox}

\subsection{RQ2: In cases where models trained on syntactical representations of code fail to learn the programming language's syntax, do they learn abstract patterns based on syntax?} \label{subsec:experiment_results:rq2}
    Our findings in RQ1 indicate that the models under study are incapable of representing the complete syntax of the programming language. However, as they are trained on syntactically valid representations of code, and as our results for cosine similarity of clone and non-clone codes across the extracted embeddings show some retention of syntax information, we aim to investigate what representation from the syntax these models learn. Furthermore, as our results for FuncGNN in RQ1 reveal what it learns and which parts of the CFGs it pays attention to for CCD, we do not further analyze it in this RQ. Therefore, to investigate the representations learned by the models, we keep the probing approach the same while changing the $<d, c, u>$ tuple construction as follows, to probe the models for abstractions of syntactic information: 
    
    \begin{itemize}
        \item $c$: A binary flag for each node that indicates whether the node has a child in the AST or not. Here, as opposed to the approach taken in RQ1, we do not include the child information for each node to be re-constructed and instead only investigate whether the models under study are capable of learning whether a node in the AST has children or not.
        \item $u$: The direct label of the node extracted from the AST. As opposed to the original $u$ tuple which was extracted from the vector representation used by each model.
        \item $d$: We have decided to discard this vector representation since the new C and U tuples, are insufficient for reconstructing the AST. As such, encoding the position information is no longer helpful.
    \end{itemize}

    Listing \ref{listing:cu_example} shows the \cu tuple constructed from the running example presented in Figure \ref{fig:dcp_annotated_ast}. 
    
    \begin{lstlisting}[float=!b, caption=Generated \cu tuple for the AST in Figure \ref{fig:dcp_annotated_ast}, label={listing:cu_example}]
        c = [ 
                1 # node 1 has children.
                1 # node 2 has children,
                1 # node 3 has children,
                .......
                0 # node 8 has no children
                0 # node 9 has no children
                .....
            ]
        u = [Module, FunctionDef, Expr, Arguments, ....]
    \end{lstlisting}
    
    \begin{table}[h]
        \caption{Result of DeepCodeProbe's accuracy on recovering Syntactic Information}
        \resizebox{0.85\textwidth}{!}{
            \centering
            \begin{tabular}{c c c r r}
                \toprule
                Model & Task & Programming Language &  Accuracy-C(\%) & Accuracy-U(\%)\\
                \midrule
                    \multirow{2}{*}{AST-NN} & \multirow{2}{*}{CCD} & C & 99.17 & 62.11 \\
                    &  & Java  & 97.50 & 61.25\\
                \midrule
                    SummarizationTF & Code Summarization & Python & 73.64 & 60.97\\
                \midrule
                    CodeSumDRL & Code Summarization & Python & 43.21 & 32.03\\
                \bottomrule
            \end{tabular}}
        
        \label{table:experiment_results::simpler_probes}
    \end{table}    
    
    Table \ref{table:experiment_results::simpler_probes} shows the results of our probing approach with the new \cu tuple. By designing a more abstract data representation, the accuracy of the probe improves significantly compared to probing for syntax for AST-NN and SummarizationTF while only a relatively small improvement is observed for CodeSumDRL. This is an important finding as it demonstrates that, for software maintenance tasks, selecting a task-specific data representation that aligns with the models' capabilities eliminates the necessity for the model to fully learn the programming language syntax; an abstraction of it suffices.
    
    Conversely, the slight enhancement observed in CodeSumDRL's probe results can be attributed to two factors:
    \begin{itemize}
        \item Even though CodeSumDRL uses ASTs to represent code, in contrast to AST-NN and SummarizationTF, it follows a multi-step process to encode information (an LSTM for processing the natural language descriptions and a Tree-based-RNN for processing the AST) as described in Section \ref{subsubsec:experiment_design:model_under_study:code_sum_drl}. As such, we will further investigate CodeSumDRL's syntax learning capabilities in RQ3 (Section \ref{subsec:experiment_results:rq3}).
        \item The complex data representation scheme which abstracts many of the details from the AST alongside the limited capacity of the model, results in loss of syntactic information.
    \end{itemize}
    
    Despite CodeSumDRL's limited capacity to represent detailed and abstract syntactic information, it achieves commendable results on established code summarization benchmarks \cite{wan2018improving}. The performance of CodeSumDRL on code summarization benchmarks alongside the high level of abstract syntax representation retention on AST-NN and SummarizationTF, indicate the importance of using explicit information encoding for small models. This observation is particularly relevant given our choice of models to study. As described in Section~\ref{subsec:experiment_design:model_under_study}, we chose models that are trained not on raw code (which contains syntax information implicitly) but on the syntactically valid representation of code (which contains the syntax explicitly). By cross-analyzing the results of RQ1 and RQ2 we can infer the following:
    \begin{itemize}
        \item For smaller ML models, using representations that explicitly encode the syntax of the programming language, allows for some level of syntax learning capability regardless of the task and models' capacity.

        \item AST and CFG representations of code exclude redundant details such as comments, docstrings, and variable names, resulting in a significantly reduced representation space compared to using raw code text as input for models. This streamlined representation space is advantageous as it allows models to learn abstractions of the programming language's syntax even when the models are not large (i.e. have a high number of learnable parameters).

    \end{itemize}

    \begin{tcolorbox}[colback=blue!5,colframe=blue!40!black]
        \textbf{Findings 2:} By using a more abstract data representation for probing the models, we can observe a significant increase in the syntactic information captured in 
        their latent space, except for CodeSumDRL which uses complex representations of ASTs. For AST-NN and SummarizationTF, we can observe that they are capable of retaining information about structural information and the relationship between nodes in an AST to accomplish their tasks without the need to learn the entire syntax of the programming language.
    \end{tcolorbox}

\subsection{RQ3: Does increasing the capacity of models without changing their architecture improve syntax learning capabilities?} \label{subsec:experiment_results:rq3}
    The findings from RQ1 and RQ2 suggest that the models studied do not fully capture the syntax of the programming language within their latent space. Instead, they tend to abstract the syntax to some degree. To explore this further, we investigate whether increasing each model's capacity—specifically, by increasing the number of parameters in the layers responsible for extracting embeddings (as outlined in Section \ref{subsec:experiment_design:scaling_up_the_models})—can enhance their learning capabilities. This investigation allows us to discern whether the models' limitation in representing full syntax stems from insufficient capacity or from inherent architectural constraints.

    Table \ref{table:experiment_results::bigger_models} presents the results of our experiments with the ``Size'' column indicating the number of parameters for each model. We use the same \dcu tuple construction and probing approach from RQ1 to probe the models under study.
    
    \begin{table}[h]
        \caption{Result of \dcp's \dcu accuracy on models with higher capacity}
        \resizebox{0.85\textwidth}{!}{
            \centering
            \begin{tabular}{c c c r r r}
                \toprule
                Model & Size & Programming Language & Accuracy-D(\%) & Accuracy-C(\%) & Accuracy-U(\%)\\
                \midrule
                    \multirow{4}{*}{AST-NN} & \multirow{2}{*}{256} & C & 13.11 & 15.02 & 15.40 \\
                    &  & Java  & 10.47 & 9.30 & 8.92 \\
                    \cline{3-6}
                    & \multirow{2}{*}{512} & C & 8.62 & 8.73 & 8.73 \\
                    &  & Java & 8.94 & 8.82 & 8.95 \\
                \midrule
                    FuncGNN & 1024 & Java & 44.24 & 98.60 & 39.90 \\
                \midrule
                    \multirow{2}{*}{SummarizationTF} & 1024 & \multirow{2}{*}{Java }& 13.82 & 13.81 & 13.74 \\
                    &  2048 &  & 13.79 & 13.60 & 13.44 \\
                \midrule
                    \multirow{3}{*}{CodeSumDRL} & 1024 & \multirow{3}{*}{Python} & 50.41 & 42.75 & 36.72\\
                    &  2048 &  & 50.23 & 42.97 & 36.30 \\
                    &  4096 &  & 43.67 & 40.22 & 34.31 \\
                    
                \bottomrule
            \end{tabular}}
        
        \label{table:experiment_results::bigger_models}
    \end{table}

    Analysing the results presented in Table \ref{table:experiment_results::bigger_models}, we can observe that:
    
    \begin{itemize}
        \item AST-NN: For both C and Java, when we scale up from the original 128 parameters to 256, we obtain an increase in the accuracy rates for $d$, $c$, and $u$ predictions. However, as we expand the model beyond this point, the rate of improvement in accuracy diminishes. This suggests that the model's capacity to capture syntactic information does not proportionally increase with size, given the architecture in place.
        
        \item FuncGNN: Scaling the number of parameters for FuncGNN from 512 to 1024, results in no noticeable difference in the accuracy of retrieving the \dcu tuples. Considering the results obtained in RQ1, where we observed that FuncGNN focuses on the edges of CFGs, by scaling up the model we do not observe any differences in the model's capability to discern between node position and node types for clone detection. Furthermore, we observe a decrease in the model's capability in CCD indicating overfitting, meaning that the model has reached its optimal capacity for the current architecture.

        \item SummarizationTF: Scaling the number of parameters from 512 to 1024, displays no improvement in increasing the accuracy of retrieving the $d$ tuple and a decrease in the probe's capability to predict the $c$ and $u$ tuples. This suggests that the model has reached its learning capacity within the current architectural constraints and scaling it up further will only result in overfitting and diminishing returns.

        \item CodeSumDRL: Increasing the number of parameters from 512 to 1024 shows a noticeable improvement in the probe's accuracy in predicting the \dcu tuples. Further scaling to 2048 parameters does not result in more improvement. However, no overfitting is observed. By further scaling to 4096 parameters, there's a drop in the probe's accuracy in predicting the \dcu tuples compared to 2048 parameters, suggesting that the model has reached its optimal capacity for the current architecture, leading to overfitting in learning syntactic representations.

    \end{itemize}

    Given the results of scaling the models, and given the architecture of each model as described in section \ref{subsec:experiment_design:model_under_study}, we can observe that increasing the models' size provides diminishing returns for each model after a certain point. Consistent with existing literature \cite{lecun2015deep}, each successive layer in a DNN learns more complex representations from the previous layers' outputs. Our results also show that beyond a certain size and without changing the models' architecture, their capacity to learn syntactic information stops improving and shows a decline, indicating potential overfitting. Moreover, the models' effectiveness on their original tasks shows little enhancement with increased size, suggesting that the initial parameter counts were already optimal and beneficial for maintaining the model's generalization capabilities. These observations, coupled with the results obtained from probing the original models on their syntax learning ability in the previous RQs lead us to an important conclusion: \textit{if the task and data representations are selected adequately, there is no need to employ models that learn the full syntax of the programming language for software maintenance.}
    
    \begin{tcolorbox}[colback=blue!5,colframe=blue!40!black]
        \textbf{Findings 3:} Scaling models trained on code, provides marginal returns on their syntax learning capabilities without showing much improvement on the models' original tasks. Given that the models are trained on syntactical representations of code, and their much smaller size compared to LLMs, our results show that there is no need to have the models learn the full syntactic rules of the programming language as long the task and data representation are selected adequately.
    \end{tcolorbox}

\section{Related Works}\label{sec:related_works}
    Given the success of BERT \cite{devlin2018bert}, researchers have conducted numerous studies on how BERT and BERT-based models learn and represent data for downstream tasks \cite{rogers2021primer}. Although these studies primarily focus on general linguistic features rather than syntax-specific to code, the methodologies developed have been extensively applied to probe larger transformer-based models on how they learn and how they represent data in their latent space. Liu et al. \cite{liu2019linguistic} investigated the linguistic capabilities of pretrained models like ELMo, OpenAI's transformer, and BERT through seventeen probing tasks. They found that while these models perform well on most tasks, they struggle with complex linguistic challenges. Hewitt and Manning \cite{hewitt2019structural} introduced a structural probe to assess whether complete syntax trees are embedded within the linear transformations of neural network word representations (i.e. embeddings). They showed that models like ELMo and BERT encode syntax trees within their vector geometries. Miaschi et al. \cite{miaschi2020contextual} compared the linguistic knowledge in the BERT's embeddings and Word2Vec. They used probing tasks targeting various sentence-level features and found that BERT's aggregated sentence representations hold comparable linguistic knowledge to Word2Vec.

With the advent of LLMs and their emerging capabilities, there is increasing interest in understanding how these models process information and make decisions based on their inputs. In the context of software engineering, researchers have adapted probing techniques originally used to understand the linguistic features that models learn from their training data to examine what models trained on code learn from their respective datasets. This has led to the creation of novel probing tasks specifically for LLMs trained on code.  Wan et al. \cite{wan2022they} have probed pre-trained language models trained on code to investigate if they learn meaningful abstractions related to to programming constructs, control structures, and data types. They conducted structural analyses from three distinct perspectives: attention analysis, probing on the word embeddings, and syntax tree induction. Their findings indicate that attention strongly aligns with the syntax structure of code, and that pre-trained models can preserve this syntax structure in their intermediate representations. Similarly, Troshin et al. \cite{troshin2022probing} evaluated the capability of pre-trained code models, such as CodeBERT \cite{feng2020codebert} and CodeT5 \cite{wang2021codet5}, in understanding source code, through diagnostic probing tasks, moving beyond task-specific models for code processing such as code generation and summarization. They introduced a variety of probing tasks to assess models' understanding of both syntactic structures and semantic properties of code, including namespaces, data flow, and semantic equivalence. Karmakar et al. \cite{karmakar2021pre}, investigated the extent to which pre-trained code models encode various characteristics of source code by designing and employing four diagnostic probes focusing on surface-level, syntactic, structural, and semantic information of code. Their methodology involves probing BERT \cite{devlin2018bert}, CodeBERT \cite{feng2020codebert}, CodeBERTa \cite{wolf2019huggingface}, and GraphCodeBERT \cite{guo2020graphcodebert} with tasks aimed at understanding the extent to which these models capture code-specific properties. Their findings show that these models can indeed capture the syntactic and semantic layers of code with varying effectiveness. However, all the models studied struggle to accurately comprehend the structural complexity of code, particularly the interconnected paths within it.

\section{Discussion}\label{sec:discussion}
    Based on the numerous experiments we conducted while replicating the results of the models under study, as described in Section \ref{subsec:experiment_design:model_under_study}, along with the results of our probing, we identify best practices for training models for software maintenance on syntactic representations of code. We expand on each of these practices as follows.

\subsection{Efficacy of Syntactical Representations in Model Performance}
   
  Our experiment results lead us to an important conclusion: models do not need to fully learn the syntax of the programming language to perform effectively on software maintenance tasks. Our analyses show that deep learning models can extract and learn abstractions of the syntax, which can be sufficient for achieving their objectives. This ability to learn abstractions stems from training on syntactically valid representations of code, rather than treating code merely as text. Traditionally, approaches for code clone detection and code summarization have trained models on code by treating code as text \cite{yang2022survey}. However, our probing reveals that leveraging artifacts extracted from code, which remove task-irrelevant information, allows us to train smaller yet effective models. These models can learn abstractions of the programming language's syntax without sacrificing performance or inference time. Empirical studies indicate that the performance of deployed ML models is often constrained during inference, as large models require extensive processing resources and time \cite{zhang2019empirical, yang2022survey, chen2020comprehensive}. By adopting syntactically informed representations for code, we can enhance model efficiency and significantly reduce model size (compared to LLMs) without compromising predictive performance on specific downstream tasks.  
         
\subsection{Tailoring Data Representations to Task-Specific Needs}
    For each of the models under study, the training data consists of either AST/CFG (FuncGNN, SummarizationTF, CodeSumDRL) or an abstraction of it (AST-NN). However, as described in Section \ref{subsec:experiment_design:model_under_study} each model follows a different data representation scheme to further break down the complexities of AST/CFG. AST-NN breaks down the ASTs into smaller STs in order to solve the problem of overly large inputs. It also uses Word2Vec representation of node labels instead of simply going over all the unique tokens and assigning them an index. This representation allows AST-NN to achieve good performance on CCD while maintaining a compact size. In contrast, FunCGNN uses statement-level tokenization of inputs instead of word-level for representing the nodes in the CFG, resulting in a smaller search space for learning the labels of each node and enabling the model to capture finer details in CFGs to achieve CCD, as shown in Section \ref{subsec:experiment_results:rq1}. SummarizationTF and CodeSumDRL are models trained to provide natural language descriptions of code. This can be viewed as a translation task (by looking at code as one language and the natural language as the target language). In fact, most comment generation and code summarization approaches are conceptualized as language translation tasks in which each block of code is systematically mapped to a corresponding natural language description \cite{yang2022survey}. However, as described in Section \ref{subsubsec:experiment_design:model_under_study:sum_tf} and Section \ref{subsubsec:experiment_design:model_under_study:code_sum_drl}, using AST extracted from code rather than treating code as text allows SummarizationTF and CodeSumDRL to achieve good performance while being much smaller than LLMs and still capable of learning abstractions from the programming language's syntax. Therefore, we recommend that for tasks requiring models to be trained on code, practitioners should use syntactic representations such as AST or CFG instead of treating code as text.

\subsection{Enhancing Model Reliability through Interpretability Probes}
    To have reliable models, an important aspect to consider is interpretability. Exploring models' behavior through probing provides insights into their decision-making. Probing helps developers pinpoint errors, refine the model with additional training data, and improve interpretability. DeepCodeProbe displays the benefits of probing by showing how it allows understanding what the models focus on to achieve their tasks. Similar approaches allow developers to investigate specific instances where models under study make errors, facilitating a better understanding of the reasons behind such mistakes.

With the rise of LLMs, there is a tendency to train models on extensive corpora of source code, which enables these models to implicitly learn the syntax of programming languages. This approach has shown success, as various probing studies on LLMs report that these models can represent programming language syntax in their latent space  \cite{hernandez2022ast, wan2022they, troshin2022probing, karmakar2021pre}.  However, LLMs are expensive to train \cite{sharir2020cost}, require enormous amounts of training data \cite{zhao2023survey}, are prone to hallucinations \cite{zhang2023siren}, and lack interpretability \cite{zhao2024explainability}. In contrast, the models we have studied are significantly smaller and less resource-intensive. Their training is computationally inexpensive compared to LLMs, and they can be trained on readily available data without complex preprocessing. Consequently, inference on these models is also inexpensive. They are all based on either RNNs, encoder-decoder, or seq2seq architectures, which are well-studied in ML and SE literature for interpretability \cite{choudhary2022interpretation}. As our study results demonstrate, these models can learn abstractions from the syntax of programming languages to achieve their tasks effectively.

While LLMs have shown impressive capabilities in code generation and understanding, relying solely on their outputs can be problematic, especially in safety-critical domains like software maintenance. The lack of interpretability and potential for hallucinations in LLMs can introduce bugs or vulnerabilities, making them unreliable for such critical tasks \cite{tambon2024bugs}. In contrast, the smaller models we studied offer advantages such as lower computational requirements, well-understood architectures, and higher confidence in their decision-making processes and learned representations. Therefore, for software maintenance tasks, developers may find it beneficial to consider using smaller, more interpretable models rather than LLMs, given the issues described above.

\section{Threats to Validity}\label{sec:threats_to_validity}
    Threats to \textbf{internal validity.} concern factors, internal to our study, that could have influenced our results. Our data representation scheme for each model which is built upon AST-Probe, can be considered a potential threat to the validity of our research. Probing has been used in other contexts in NLP for uncovering what the models learn, and our results show that by extracting a syntactically valid vector representation, we can probe models for syntactical information. Additionally, our data transformation and embedding extraction for each of the models under study, may not encompass all the required steps to understand the internal representations of DL models. We mitigate this threat by experimenting with multiple data representation approaches and extracting the embeddings from different layers of the models under study to have a comprehensive understanding of the probing task and the obtained results.

Threats to \textbf{construct validity} concern the relationship between theory and observation. The primary threat to the construct validity of our study is the probing technique we propose. This technique could potentially provide a distorted view of what models have truly learned, thereby impacting our analysis and the conclusions drawn from it. In order to mitigate this threat, we validate the extracted embeddings by comparing the cosine similarity of code clone and non-clone pairs from models trained on syntactically valid code representations. Furthermore, we establish the probe's validity by demonstrating that it does not learn new representations but reveals pre-existing information in the model's latent space. Moreover, we have conducted thorough experiments using embeddings from both trained and randomly initialized models, ensuring our probing approach robustly reflects the underlying syntactic learning of the models.

Threats to \textbf{conclusion validity} concern the relationship between theory and outcome and are mainly related to our analysis done on the our probes' results. To mitigate this threat we have conducted probing on models with the same architecture but higher capacities to ensure that our conclusions and the suggested guidelines are based on both theoretical and empirical evidence.

Threats to \textbf{external validity} concern the generalization of our findings. We have selected models for software maintenance across two different tasks with different data representation schemes and trained on different programming languages, in order to have a comprehensive analysis and mitigate this threat. Furthermore, there exists the possibility that our probing approach produces different results than the ones reported on other models with relatively the same capacity but different architecture and data processing approaches. As there exists a plethora of ML models for software maintenance tasks, we have selected the models under study in a manner to cover most of the design choices that are being utilized (encoder-decoder models, graph neural networks, and seq2seq models). It should be noted that in order to be able to validate and analyze the results of our proposed probing approach, we were required to study models that had made both their code and data available and were trained on syntactically correct code, to replicate their results and ensure that any negative results and analysis were not resulted from incorrect replication. Given such limitations, the models under study cover the majority of the proposed approaches across programming languages and architectural designs. Given that our data transformation and embedding extraction are model agnostic, we believe that our probing approach can be used to probe other models as well. 

Threats to \textbf{reliability and validity} concern the possibility of replicating this study. We have provided all the necessary details needed for replication, sharing our full replication package~\cite{rep_package}. Additionally, since our probing approach requires re-training the models under study, our replication package includes the code, links to data repositories, and the configurations for re-training the models.

\section{Conclusion}\label{sec:conclusion}
    In this study, we introduced \dcp, a novel probing approach for assessing the syntax learning capabilities of non-transformer-based, ML models trained on code for software maintenance tasks. Our probing approach is based on syntactically valid constructs extracted from codes which allows for efficient and reliable probing of ML models on their syntax learning capabilities alongside providing interpretability on the representations learned by the models to achieve their tasks. Our results show that while non-transformer-based, ML models are unable to fully capture the complete set of syntactic rules of programming languages, they are capable of learning relevant abstractions and patterns within the language syntax. This suggests that such models, despite their relatively simple and small architectures compared to LLMs, can still acquire meaningful syntactic knowledge from code-based training data. Furthermore, our experiments demonstrate that simply increasing the capacity of these models, without changing their architecture, leads to only marginal improvements in their full syntax learning abilities. These results indicate that the architectural design of the model plays a crucial role in determining its syntax learning capabilities, beyond just the amount of training data or model size. These findings have important implications for the development of small, efficient ML models for code-related tasks, as they highlight the need to carefully consider model architecture choices in addition to scaling up model size. Finally, we provide a set of guidelines based on our proposed probing approach and the subsequent conducted analysis on the models under study for training ML models for software maintenance.

To allow for the development and deployment of more reliable ML models for software maintenance, we aim to expand \dcp~ into a framework that can be deployed on top of trained models. Doing so would enable developers and researchers to systematically evaluate and enhance the syntax learning capabilities of ML models, ensuring their effectiveness in software maintenance tasks. 

\section{Acknowledgments}\label{sec:acknowledgments}
    This work is partially supported by the Fonds de Recherche du Quebec (FRQ), the Canadian Institute for Advanced Research (CIFAR), and the Natural Sciences and Engineering Research Council of Canada (NSERC).

\bibliographystyle{ACM-Reference-Format}
\bibliography{sample-base}

\end{document}